\documentclass[twocolumn,aps,prb,cite,superscriptaddress,groupedaddress,showpacs,epsfig]{revtex4}
\usepackage{epsfig}
\begin{document}
\title{
\Large\bf Field theory of bi- and tetracritical points:  Statics}
\author{R. Folk}\email{folk@tphys.uni-linz.ac.at}
\affiliation{Institute for Theoretical Physics, Johannes Kepler
University Linz, Altenbergerstrasse 69, A-4040, Linz, Austria}
 \author{Yu. Holovatch}\email[]{hol@icmp.lviv.ua}
 \affiliation{Institute for Condensed Matter Physics, National
Academy of Sciences of Ukraine, 1~Svientsitskii Str., UA--79011
Lviv, Ukraine}\affiliation{Institute for Theoretical Physics, Johannes Kepler
University Linz, Altenbergerstrasse 69, A-4040, Linz, Austria}
  \author{G. Moser}\email[]{guenter.moser@sbg.ac.at}
\affiliation{Department for Material Research and Physics, Paris Lodron University
Salzburg, Hellbrunnerstrasse 34, A-5020 Salzburg, Austria}
\date{\today}
\begin{abstract}
We calculate the static critical behavior of systems of
$O(n_\|)\oplus O(n_\perp)$ symmetry by renormalization group method
within the minimal subtraction scheme in two loop order. Summation
methods lead to fixed points describing multicritical behavior.
Their stability boarder lines in the space of order parameter
components  $n_\|$ and $n_\perp$ and spatial dimension $d$ are
calculated.  The essential features obtained already in two loop
order for the interesting case of an antiferromagnet in a magnetic
field ($n_\|=1$, $n_\perp=2$) are the stability of the biconical
fixed point  and the neighborhood of the stability border lines to
the other fixed points leading to very small transient exponents. We
are also able to calculate the flow of static couplings, which
allows to consider the attraction region. Depending on the
nonuniversal background parameters the existence of different
multicritical behavior (bicritical or tetracritical) is possible
including a triple point.

\end{abstract}
\maketitle
\section{Introduction}

Antiferromagnets in an external magnetic field show a variety of
phase diagrams depending on the interaction terms present in the
spin Hamiltonian\cite{liufi72}. The spin interaction may be
isotropic, anisotropic with an easy axis and/or single-ion
anisotropy terms, where the anisotropy is in the direction of the
external magnetic field. The phase diagram of such models exhibit a
multicritical point, where several transition lines meet.

At a {\it bicritical} point three phases - an antiferromagnetic
phase, a spin flop phase and the paramagnetic phase - are in
coexistence. The phase transition lines to the paramagnetic phase
are second order transition lines, whereas the transition line
between the spin flop and the antiferromagnetic phase is of first
order. At the {\it tetracritical} point four phases - an
antiferromagnetic phase, a spin flop phase, an intermediate or mixed
phase and the paramagnetic phase - are in coexistence. All
transition lines are of second order in this case.

A field theoretic description of these models starts with a static
functional for an $n$-component field $\Phi$ of $O(n_\|)\oplus
O(n_\perp)$ symmetry ($n_\| + n_\perp=n$) leading to different
multicritical behavior connected with the stable fixed point (FP)
found in the renormalization group
treatment\cite{nekofi74,ahabru74,konefi76}. Bicritical behavior has
been connected with the stability of the well known {\it isotropic
Heisenberg} fixed point of $O(n_\|+n_\perp)$ symmetry, whereas
tetracriticality has been connected with a fixed point of
$O(n_\|)\oplus O(n_\perp)$ symmetry, which might be either the so
called {\it biconical} FP or the {\it decoupling} FP. In the last FP
the parallel and the perpendicular components of the order parameter
(OP) are asymptotically decoupled.

The important questions which theory should give an answer to is,
(i) which of these FPs is the stable one in a three- or
two-dimensional system, and (ii) what are the differences in the
critical behavior at the multicritical point? These questions have
been risen and considered in one loop order\cite{konefi76}, where
the Heisenberg FP turns out to be the stable one in $d=3$ for the
case $n_\|=1$ and $n_\perp=2$, but this picture is changed in higher
loop order. In a five loop order $\epsilon=4-d$ expansion it has
been found that the biconical FP is the stable one \cite{capevi03}.
It also has been found that the differences between the exponents at
the different multicritical points are much smaller than in the one
loop order calculation.

Physical examples where such multicritical behavior has been found
are the anisotropic antiferromagnets\cite{shapira83} (with the
magnetic field in the hard direction) like\cite{ro75,roge77}
GdALO$_3$ and\cite{kiro79} MnF$_2$, as well as\cite{bucohathmu81}
MnCl$_2$4D$_2$O  or\cite{ohue05} Mn$_2$AS$_4$ (A=Si or Ge). Other
examples with a single ion anisotropy  might be layered cuprate
antiferromagnets like (Ca,La)$_{14}$Cu$_{24}$O$_{41}$. Besides the
examples with $n_\|=1$ and $n_\perp=2$ one might consider other
cases: $n_\|=1$ and $n_\perp=1$ when additional anisotropies are
present as in\cite{beccera88,basten80} NiCl$_2$ or high-$T_c$
superconductors representing a system with $n_\|=2$ (corresponding
to the superconductor OP) and $n_\perp=3$ (corresponding to the
antiferromagnetic OP).

Quite recently the possible types of  phase diagrams in the magnetic
field - temperature plane of $d=3$ of uniaxially anisotropic
antiferromagnets have been studied by Monte Carlo
simulations\cite{selke05,selkeetal08}. For $n_\|=1$ and $n_\perp=2$
a phase diagram with a bicritical point has been found in agreement
with earlier simulations \cite{labi78}, but contrary to the results
of renormalization group theory in higher loop
orders\cite{capevi03}.

A general picture is obtained when one considers a generalized model
with an $n$ component order parameter (OP), which splits into $n_\|$
parallel OP components and $n_\perp$ perpendicular OP components and
quartic interaction terms of $O(n_\|)\oplus O(n_\perp)$ symmetry.
Both parallel and perpendicular OP components become critical at the
multicritical point. In the $n_\|$-$n_\perp$-space regions of
different types of multicriticality exist touching each other at
stability border lines ('phase border lines') where the fixed points
change their stability (such a picture of the different stability
regions might be called a 'phase diagram').  In addition to the
stability of a fixed point we want to mention that one has to
consider also the attraction regions of a fixed point to answer the
question wether one can reach the stable fixed point. In order to
discuss the attraction regions one has to consider the flow of the
couplings from the nonuniversal initial (background) values.

We therefore reconsider the critical behavior of systems with
$O(n_\|)\oplus O(n_\perp)$ symmetry. Being interested in criticality
of three dimensional systems, we will work  within the minimal
subtraction scheme and evaluate the results at  fixed dimension
$d=3$.\cite{Schloms} For the  universal properties (as asymptotic
critical exponents and marginal dimensions) it turns out that
already the two loop calculations refined by resummation are in good
quantitative agreement with previous resummed higher order
$\varepsilon$-expansion results.\cite{capevi03} However, contrary to
previous calculations, the technique we use gives a possibility to
analyze non-universal effective critical behavior which is
manifested in a broader temperature interval near the
(multi)critical point. Such calculations are out-of-reach the
$\varepsilon$-expansion and will be performed below on the base of
analysis of the renormalization group flow.

The paper is organized as follows: starting from the static
functional (Sec. \ref{func}) we introduce the renormalization in
Sec. \ref{renorm} and calculate the field theoretic functions in
Sec. \ref{zeta}. The perturbative expansions being asymptotic, we
apply in Sec. \ref{fixp} the resummation technique to restore their
convergence and to extract numerical values of the fixed points of
the renormalization group transformations and their stability. We
discuss the stability border lines between the fixed points and show
that they are shifted considerably compared to the one loop
calculation. As a result for the isotropic antiferromagnet in a
magnetic field represented by the point (1,2) in the
$n_\|$-$n_\perp$-space the {\it biconical fixed point is stable
predicting tetracritical behavior if the fixed point is reached from
the background}. The very neighborhood of  the stability border
lines is characterized by very small transient exponents. Moreover,
looking at the attraction regions a surface in the space of the
fourth order couplings exists above which {\it no finite fixed
point} can be reached. This indicates the possibility of a scenario
mentioned already earlier\cite{domufi77,muna00,capevi03} where the
multicritical point is a {\it triple point} and the second order
lines separating the paramagnetic phase from the ordered phases
contain a tricritical point. In section \ref{expon} the critical
exponents are defined and their asymptotic values are calculated for
the physically interesting case $n_\|=1$, $n_\perp=2$.  The flow
equations and effective exponents are discussed in Sec. \ref{flow}
leading to our final conclusions and outlook in Sec. \ref{out}. In
the appendices we discuss the perturbative expansion for the vertex
functions (Appendix A) and explain the resummation procedure
exploited in our calculations (Appendix B).

\section{Static functional} \label{func}

The critical behavior of an isotropic system ($O(n)$ symmetry) with short range
interaction is determined by the static functional
\begin{eqnarray}\label{hglw}
{\cal H}_{GLW}\!=\!\int\! d^dx\Bigg\{\frac{1}{2}\mathring{r}\vec{\phi}_0
\cdot\vec{\phi}_0+\frac{1}{2}\sum_{i=1}^n\nabla_i\vec{\phi}_0\cdot
\nabla_i\vec{\phi}_0  \nonumber \\
+\frac{\mathring{u}}{4!}\Big(\vec{\phi}_0\cdot\vec{\phi}_0\Big)^2 \Bigg\} \ ,
\end{eqnarray}
which is known as the Ginzburg-Landau-Wilson(GLW)-functional. The
order parameter $\vec{\phi}_0\equiv \vec{\phi}_0(x)$ is assumed to
be a $n$-component real vector. The symbol $\cdot$ denotes the
scalar product between vectors. $\mathring{r}$ is proportional to
the temperature distance to the critical point and $\mathring{u}$ is
the fourth order coupling in which perturbation expansion is usually
performed. Systems represented by such a static functional have been
extensively studied in the last decades with different
renormalization procedures, and the corresponding critical exponents
and amplitude ratios are known up to high loop orders (see
e.g.\cite{rgbooks}).

In order to describe bicritical behavior the $n$-dimensional space of the order
parameter components will
be divided into two subspaces with dimensions $n_\perp$ and $n_\|$ with the
property $n_\perp+n_\|=n$ in the following. Correspondingly the order parameter
separates into
\begin{equation}\label{phisplit}
\vec{\phi}_0=\left(\begin{array}{c} \vec{\phi}_{\perp 0} \\  \vec{\phi}_{\| 0}
\end{array}\right)
\end{equation}
where $\vec{\phi}_{\perp 0}$ is the $n_\perp$-dimensional order
parameter of the $n_\perp$-subspace, and $\vec{\phi}_{\| 0}$ is the
$n_\|$-dimensional order parameter of the $n_\|$-subspace.
Performing the separation in the GLW-functional (\ref{hglw}) one
obtains

\begin{eqnarray}\label{hbicrit}
{\cal H}_{Bi}\!=\!\int\! d^dx\Bigg\{\frac{1}{2}\mathring{r}_\perp\vec{\phi}_{\perp 0}
\cdot\vec{\phi}_{\perp 0}+\frac{1}{2}\sum_{i=1}^{n_\perp}\nabla_i\vec{\phi}_{\perp 0}\cdot
\nabla_i\vec{\phi}_{\perp 0}  \nonumber \\
+\frac{1}{2}\mathring{r}_\|\vec{\phi}_{\| 0}
\cdot\vec{\phi}_{\| 0}+\frac{1}{2}\sum_{i=1}^{n_\|}\nabla_i\vec{\phi}_{\| 0}\cdot
\nabla_i\vec{\phi}_{\| 0}  \nonumber \\
+\frac{\mathring{u}_\perp}{4!}\Big(\vec{\phi}_{\perp 0}\cdot\vec{\phi}_{\perp 0}\Big)^2
+\frac{\mathring{u}_\|}{4!}\Big(\vec{\phi}_{\| 0}\cdot\vec{\phi}_{\| 0}\Big)^2
\nonumber \\
+\frac{2\mathring{u}_\times}{4!}\Big(\vec{\phi}_{\perp
0}\cdot\vec{\phi}_{\perp 0}\Big) \Big(\vec{\phi}_{\|
0}\cdot\vec{\phi}_{\| 0}\Big) \Bigg\} \ .
\end{eqnarray}
This functional contains three fourth order couplings
$\{\mathring{u}\}=\{\mathring{u}_\perp,\mathring{u}_\times,\mathring{u}_\|\}$
and instead of one parameter $\mathring{r}$ as in (\ref{hglw}), in
(\ref{hbicrit}) two different parameters $\mathring{r}_\perp$ and
$\mathring{r}_\|$ appear referring to different temperature
distances.

The decomposition in parallel and perpendicular OP components allows
to describe the critical behavior at the meeting point of two
critical lines: (i) the line where $\mathring{r}_\perp$ becomes zero
and the $n_\perp$-dimensional components $\vec{\phi}_{\perp 0}$ are
the OP, and (ii) the line where $\mathring{r}_\|$ becomes zero and
the $n_\|$-dimensional components $\vec{\phi}_{\|0}$ are the OP. At
the meeting point both quadratic terms become zero and both
components of  $\vec{\phi}_{0}$ have to be taken into account. The
critical behavior of this multicritical point, has been described
already in one loop order\cite{konefi76} and three different types
of multicritical behavior have been found; (i) one described by the
well known isotropic $n$ component Heisenberg fixed point where all
fourth order couplings are equal, (ii) one described by a decoupling
fixed point, which consists of a combination of two $n_\perp$ and
$n_\|$ component isotropic Heisenberg fixed points of two decoupled
systems and (iii) a new type of fixed point called the biconical
fixed point. Which of these fixed point is the stable one depends on
the number of components $n_\perp$ and $n_\|$ and the dimension $d$
of space. The scaling properties depend on the symmetry of stable
fixed point When the  $O(n)$ symmetry in the OP space is broken to
$O(n_\|)\oplus O(n_\perp)$ symmetry, then also the spatial
correlations are different for the two OP subspaces.

\section{Renormalization} \label{renorm}

The procedure used to obtain the vertex functions appropriate for
the renormalization is described in more details in the Appendix
\ref{vertex}. From the general structure of the two point vertex
functions presented therein follows that the order parameter
functions $\vec{\phi}_{\perp 0}$ and $\vec{\phi}_{\| 0}$ in the
subspaces may be renormalized by the scalar renormalization factors
\begin{equation}\label{phiren}
\vec{\phi}_{\perp 0}=Z_{\phi_\perp}^{1/2}\vec{\phi}_{\perp} \ ,
\qquad \vec{\phi}_{\| 0}=Z_{\phi_\|}^{1/2}\vec{\phi}_{\|}.
\end{equation}
The above relations and the definitions (\ref{correlperp}) and
(\ref{correlpara}) imply that the correlation lengths, $\xi_\|$ and
$\xi_\perp$, do not renormalize. They constitute together with the
wave vector modulus $k$ the independent lengths of the system. The
fourth order couplings may also be renormalized by scalar
renormalization factors
\begin{eqnarray}
\label{uperpren}
\mathring{u}_\perp&=&\kappa^\varepsilon Z_{\phi_\perp}^{-2}Z_{u_\perp}u_\perp A_d^{-1}, \\
\label{utimesren}
\mathring{u}_\times&=&\kappa^\varepsilon Z_{\phi_\perp}^{-1}Z_{\phi_\|}^{-1}Z_{u_\times}
u_\times A_d^{-1}, \\
\label{upararen} \mathring{u}_\|&=&\kappa^\varepsilon
Z_{\phi_\|}^{-2}Z_{u_\|}u_\|A_d^{-1},
\end{eqnarray}
with the geometrical factor
\begin{equation}\label{ad}
A_d=\Gamma\!\left(1-\frac{\epsilon}{2}\right)\Gamma\!\left(1+\frac{\epsilon}{2}\right)
\frac{\Omega_d}{(2\pi)^d}
\end{equation}
 analogously to Dohm\cite{dohmKFA}. In (\ref{ad}) $\Gamma(x)$ denotes the Euler Gamma-function and
$\Omega_d$ is the surface of the $d$-dimensional unit sphere. We want to remark
that it would be possible to introduce a $3\times 3$-matrix for the renormalization of the three fourth order couplings
$\{\mathring{u}\}$ as performed in \cite{capevi03}. The resulting $\beta$-functions
are the same anyway.
Thus it remains a matter of taste if one uses a scalar or a matrix renormalization for
the three fourth order couplings. The situation changes if one considers $\phi^2$-
insertions. The vertex function $\mathring{\Gamma}^{(2,1)}$ splits up into four
functions $\mathring{\Gamma}^{(2,1)}_{\perp\perp;\perp}$,
$\mathring{\Gamma}^{(2,1)}_{\perp\perp;\|}$, $\mathring{\Gamma}^{(2,1)}_{\|\|;\perp}$
and $\mathring{\Gamma}^{(2,1)}_{\|\|;\|}$ (for the notation see Appendix \ref{vertex}).
A consistent renormalization of the $\phi^2$- insertions is only possible by introducing
\begin{equation}\label{phi2ren}
\left(\begin{array}{c} \vec{\phi}_{\perp}^2 \\  \vec{\phi}_{\|}^2
\end{array}\right)=\mbox{\boldmath$Z$}_{\phi^2}
\left(\begin{array}{c} \vec{\phi}_{\perp 0}^2 \\  \vec{\phi}_{\| 0}^2
\end{array}\right)
\end{equation}
with the renormalization matrix
\begin{equation}\label{zphi2}
\mbox{\boldmath$Z$}_{\phi^2}=\left(\begin{array}{cc} Z_{11} & Y_{12} \\  Y_{21} & Z_{22}
\end{array}\right)\, .
\end{equation}
The zeroth order of the perturbation expansion only appears in the
diagonal elements $Z_{ii}=1+{\cal O}(\{u\})$, while the off-diagonal
elements $Y_{ij}={\cal O}(\{u\})$ start with the one loop order.
With (\ref{phiren}) and (\ref{phi2ren}) an arbitrary  vertex
function (\ref{vertexfunc3}) renormalizes as
\begin{eqnarray}\label{vertexren}
\Gamma^{(N,L)}_{\alpha_1\cdots\alpha_N;\beta_1\cdots\beta_L}=
Z_{\phi_{\alpha_1}}^{1/2}\cdots Z_{\phi_{\alpha_N}}^{1/2}  \nonumber \\
\times\sum_{i_1,\cdots ,i_L}
(\mbox{\boldmath$Z$}_{\phi^2})_{\beta_1i_1}\cdots(\mbox{\boldmath$Z$}_{\phi^2})_{\beta_Li_L}
\mathring{\Gamma}^{(N,L)}_{\alpha_1\cdots\alpha_N;i_1\cdots i_L}
\end{eqnarray}
The indices $\alpha_j$, $\beta_j$ and $i_j$ are running over $\perp$
and $\|$. The renormalization introduced in the above relations
removes the poles from all vertex functions except the functions
$\Gamma^{(0,2)}_{;\beta_1\beta_2}$ which have to be considered
separately. Thus equation (\ref{vertexren}) is valid for all vertex
functions except $\Gamma^{(0,2)}_{;\beta_1\beta_2}$. The
multiplicative renormalization in (\ref{vertexren}) does not remove
all singularities at $d=4$ in these function. In order to remove the
remaining poles an additional additive renormalization is necessary.
The multiplicative renormalization leads to functions
\begin{eqnarray}\label{gamma02r}
\Gamma^{(0,2)}_{R;\beta_1\beta_2}= \sum_{i_1,i_2}
(\mbox{\boldmath$Z$}_{\phi^2})_{\beta_1i_1}(\mbox{\boldmath$Z$}_{\phi^2})_{\beta_2i_2}
\mathring{\Gamma}^{(0,2)}_{;i_1 i_2} \, .
\end{eqnarray}
A finite function can be obtained by subtracting the singular part
$[\Gamma^{(0,2)}_{R;\beta_1\beta_2}]_S$ from (\ref{gamma02r}). Thus we may introduce
\begin{equation}\label{gamma02ren}
\Gamma^{(0,2)}_{;\beta_1\beta_2}=\kappa^\varepsilon A_d^{-1}
\left(\Gamma^{(0,2)}_{R;\beta_1\beta_2}-[\Gamma^{(0,2)}_{R;\beta_1\beta_2}]_S\right)
\end{equation}
as renormalized functions. The additive renormalizations are then defined by
\begin{equation}\label{au}
A_{\beta_1\beta_2}(\{u\})=\kappa^\varepsilon A_d^{-1}
[\Gamma^{(0,2)}_{R;\beta_1\beta_2}]_S
\end{equation}
The three functions $A_{\perp\perp}$, $A_{\|\|}$ and
$A_{\perp\|}=A_{\|\perp}$ in (\ref{au}) represent the extension of
the  function $A(u)$ in the isotropic case \cite{fomorev} and may be
written as the symmetric matrix
\begin{equation}\label{amatrix}
\mbox{\boldmath$A$}(\{u\})=\left(\begin{array}{cc} A_{\perp\perp} &
A_{\perp\|} \\ A_{\perp\|} & A_{\|\|}\end{array}\right) \, .
\end{equation}

Within statics it may also be convenient to work with the
temperature dependent vertex functions  without introducing the
correlation length, as described in the Appendix at stage
(\ref{vertexfunc2}). This allows to avoid functions with
$\phi^2$-insertions except for the specific heat. In this case a
renormalization for the temperature distances
$\bigtriangleup\mathring{r}_\perp$ and
$\bigtriangleup\mathring{r}_\|$ has to be introduced as performed in
\cite{dohmKFA}. In order to obtain renormalization factors which do
not contain ratios of the temperature distances a matrix
renormalization
\begin{equation}\label{deltarren}
\bigtriangleup\mathring{\vec{r}}\equiv
\left(\begin{array}{c} \bigtriangleup\mathring{r}_\perp \\  \bigtriangleup\mathring{r}_\|
\end{array}\right)=\mbox{\boldmath$Z$}_{\phi}^{-1}\cdot\mbox{\boldmath$Z$}_r
\cdot\bigtriangleup\vec{r}
\end{equation}
with the matrices
\begin{equation}\label{zrzphi}
\mbox{\boldmath$Z$}_r=\left(\begin{array}{cc} Z_{r_\perp} & Y_{r_\perp} \\
Y_{r_\|} & Z_{r_\|}\end{array}\right) \ , \qquad
\mbox{\boldmath$Z$}_\phi=\left(\begin{array}{cc} Z_{\phi_\perp} & 0 \\
0 & Z_{\phi_\|}\end{array}\right)
\end{equation}
has to be used. The renormalization factors in the matrix above are obtained by
collecting the $\varepsilon$-poles proportional to $\bigtriangleup r_\perp$ and
$\bigtriangleup r_\|$ in the two vertex functions
$\mathring{\Gamma}^{(2,0)}_{\perp\perp}$ and $\mathring{\Gamma}^{(2,0)}_{\|\|}$.
The matrices in (\ref{zrzphi}) are related to the renormalization
matrix $\mbox{\boldmath$Z$}_{\phi^2}$ for the $\phi^2$-insertions by
\begin{equation}\label{zrzphi2rel}
\mbox{\boldmath$Z$}_{\phi^2}=(\mbox{\boldmath$Z$}_\phi^{-1}\mbox{\boldmath$Z$}_r)^T
=\mbox{\boldmath$Z$}_r^T\mbox{\boldmath$Z$}_\phi^{-1}
\end{equation}
which represents the matrix counterpart of the scalar relation in the isotropic case.
The superscript $^T$ denotes the transposed matrix.

\section{$\zeta$ - and $\beta$ - functions} \label{zeta}

From the scalar renormalization factors $Z_{\phi_{\alpha_i}}$ ($\alpha_i=\perp,\|$)
the $\zeta$-functions
\begin{equation}\label{zetafu}
\zeta_{\phi_{\alpha_i}}(\{u\})=\frac{d\ln Z_{\phi_{\alpha_i}}^{-1}}{d\ln\kappa}
\end{equation}
are derived. The $\kappa$-derivatives, also in the following
definitions, always are taken at fixed unrenormalized parameters.
In two loop order we obtain the $\zeta$-functions
\begin{eqnarray}
\label{zetaphiperp}
\zeta_{\phi_{\perp}}&=& -\frac{n_\perp+2}{72}u_\perp^2-\frac{n_\|}{72}u_\times^2 \\
\label{zetaphipara}
\zeta_{\phi_{\|}} &=& -\frac{n_\perp}{72}u_\times^2 -\frac{n_\|+2}{72}u_\|^2
\end{eqnarray}
From the matrix $\mbox{\boldmath$Z$}_{\phi^2}$ of the
renormalization of the $\phi^2$-insertions it is convenient to introduce the
$\zeta$-matrix
\begin{equation}\label{zetaphiq}
[\mbox{\boldmath$\zeta$}_{\phi^2}]_{ij}(\{u\})=-\sum_l
\left(\kappa\frac{d}{d\kappa}[\mbox{\boldmath$Z$}_{\phi^2}]_{il}\right)
[\mbox{\boldmath$Z$}_{\phi^2}^{-1}]_{lj}
\end{equation}
A matrix $\mbox{\boldmath$\zeta$}_r$ following from $\mbox{\boldmath$Z$}_r$, which
has been defined in (\ref{zrzphi}), also can be introduced analogous to
(\ref{zetaphiq}) by replacing $\mbox{\boldmath$Z$}_{\phi^2}$ with
$\mbox{\boldmath$Z$}_r$. In two loop order the components of the matrix
$\mbox{\boldmath$\zeta$}_{\phi^2}$ are
\begin{eqnarray}
\label{zetaps11}
[\mbox{\boldmath$\zeta$}_{\phi^2}]_{11}&=& \frac{n_\perp+2}{6}u_\perp
\left(1-\frac{5}{12}u_\perp\right)-\frac{n_\|}{72}u_\times^2  \\
\label{zetaps12}
[\mbox{\boldmath$\zeta$}_{\phi^2}]_{12}&=& \frac{n_\perp}{6}u_\times
\left(1-\frac{u_\times}{3}\right)
\\
\label{zetaps21}
[\mbox{\boldmath$\zeta$}_{\phi^2}]_{21}&=& \frac{n_\|}{6}u_\times
\left(1-\frac{u_\times}{3}\right)
\\
\label{zetaps22}
[\mbox{\boldmath$\zeta$}_{\phi^2}]_{22} &=& \frac{n_\|+2}{6}u_\|
\left(1-\frac{5}{12}u_\|\right)-\frac{n_\perp}{72}u_\times^2
\end{eqnarray}
The non diagonal elements (\ref{zetaps12}) and (\ref{zetaps21}) are proportional
to $u_\times$ and differ only in a prefactor $n_\perp$ and $n_\|$ respectively.
Thus we may write
\begin{equation}\label{zetaphindrel}
[\mbox{\boldmath$\zeta$}_{\phi^2}]_{12}=n_\perp u_\times C(\{u\})=
\frac{n_\perp}{n_\|}[\mbox{\boldmath$\zeta$}_{\phi^2}]_{21}
\end{equation}
with a function
\begin{equation}\label{cu}
C(\{u\})=\frac{1}
{6}(1-\frac{u_\times}{3})
\end{equation}
in two loop order. Relation (\ref{zrzphi2rel}) between $\mbox{\boldmath$Z$}_{\phi^2}$
and $\mbox{\boldmath$Z$}_r$ implies
\begin{equation}\label{zetarel}
\mbox{\boldmath$\zeta$}_{\phi^2}^T=\mbox{\boldmath$\zeta$}_r-\mbox{\boldmath$\zeta$}_{\phi}
\end{equation}
where the diagonal matrix $\mbox{\boldmath$\zeta$}_{\phi}
=\mbox{diag}(\zeta_{\phi_\perp},\zeta_{\phi_\|})$ has been introduced.
From the additive renormalization (\ref{amatrix}) the function
\begin{eqnarray}\label{bphi2}
[\mbox{\boldmath$B$}_{\phi^2}]_{ij}(\{u\})=\kappa^\varepsilon
\sum_{n,p}[\mbox{\boldmath$Z$}_{\phi^2}]_{in}[\mbox{\boldmath$Z$}_{\phi^2}]_{jp}
 \nonumber \\
\times\sum_{l,m}\kappa\frac{d}{d\kappa}\kappa^{-\varepsilon}
[\mbox{\boldmath$Z$}_{\phi^2}^{-1}]_{nl}[\mbox{\boldmath$Z$}_{\phi^2}^{-1}]_{pm}
[\mbox{\boldmath$A$}]_{lm}(\{u\})
\end{eqnarray}
can be introduced which is the extension of the scalar function $B_{\phi^2}(u)$ in
the isotropic case (see Ref.\cite{fomorev} for definitions). Calculating (\ref{bphi2})
in two loop order we obtain
\begin{equation}\label{bphi2twoloop}
\mbox{\boldmath$B$}_{\phi^2}(\{u\})=\left(\begin{array}{cc}
\frac{n_\perp}{2} & 0 \\ 0 & \frac{n_\|}{2}\end{array}\right)
+\mbox{\boldmath${\cal O}$}(\{u^2\})
\end{equation}
The $\beta$-functions of the four couplings are defined as
\begin{equation}\label{betaua}
\beta_{u_a}(\{u\})=\kappa\frac{du_a}{d\kappa}
\end{equation}
with $a=\perp,\|,\times$.  In two loop order the explicit
expressions of the $\beta$-functions are
\begin{eqnarray}
\label{betauperp2l}
\beta_{u_\perp}&=& -\varepsilon  u_\perp + \frac{(n_\perp+8)}{6} u_\perp^2
+ \frac{n_\|}{6}u_\times^2 \nonumber \\
&-& \frac{(3n_\perp + 14)}{12} u_\perp^3 - \frac{5n_\|}{36}u_\perp u_\times^2
- \frac{n_\|}{9}u_\times^3;\\
\label{betautimes2l}
\beta_{u_\times} & = & -\varepsilon  u_\times + \frac{(n_\perp+2)}{6}u_\perp u_\times +
\frac{(n_\|+2)}{6}u_\times u_\| \nonumber  \\
&+&\frac{2}{3}u_\times^2 - \frac{(n_\perp+n_\|+16)}{72} u_\times^3
- \frac{(n_\perp+2)}{6}u_\times^2 u_\perp \nonumber \\
&-& \frac{(n_\|+2)}{6} u_\times^2  u_\| - \frac{5(n_\perp+2)}{72}
 u_\perp^2 u_\times  \nonumber \\
&-& \frac{5(n_ \|+2)}{72}u_\times u_\|^2; \\
\label{betaupara2l}
\beta_{u_\|} &=& -\varepsilon  u_\| + \frac{(n_\|+8)}{6}  u_\|^2 +
\frac{n_\perp}{6}u_\times^2 \nonumber  \\
&-& \frac{(3n_\| + 14)}{12} u_\|^3 -\frac{5n_\perp}{36}u_\|u_\times^2 - \frac{n_\perp}{9}u_\times^3.
\end{eqnarray}
The flow equations of the fourth order couplings $u_a$ are
\begin{equation}
\label{duadl}
l\frac{du_a}{dl}=\beta_{u_a}(\{u\})  \\
\end{equation}
where $l$ is the flow parameter.

\begin{widetext}

\begin{table}[h]
\centering \tabcolsep=5mm
\begin{tabular}{lllllll}
 \hline \hline
   FP & $u_\perp^\star$ & $u_\times^\star$ & $u_\|^\star$ & $\omega_1$ & $\omega_2$ & $\omega_3$
   \\ \hline
   ${\mathcal G}$ & 0 & 0 & 0 & $-\varepsilon$ & $-\varepsilon$ & $-\varepsilon$
   \\
   $\mathcal{H}(n_\perp)$ & $u^{\mathcal{H}(n_\perp)}$ & 0 & 0 & $\omega^{\mathcal{H}(n_\perp)}$
& $\omega_2^{\mathcal{H}(n_\perp)}$ & $-\varepsilon$
    \\
$\mathcal{H}(n_\|)$ & 0 &  0 & $u^{\mathcal{H}(n_\|)}$ &  $-\varepsilon$ & $\omega_2^{\mathcal{H}(n_\|)}$
& $\omega^{\mathcal{H}(n_\|)}$
    \\
    $\mathcal{D}$ & $u^{\mathcal{H}(n_\perp)}$ &  0 & $u^{\mathcal{H}(n_\|)}$ &  $\omega^{\mathcal{H}(n_\perp)}$
&  $\omega_2^{\mathcal D}$ & $\omega^{\mathcal{H}(n_\|)}$
    \\
    $\mathcal{H}(n_\perp+n_\|)$ & $u^{\mathcal{H}(n_\perp+n_\|)}$ &  $u^{\mathcal{H}(n_\perp+n_\|)}$ &
$u^{\mathcal{H}(n_\perp+n_\|)}$ &  $\omega_1^{\mathcal{H}(n_\perp+n_\|)}$
&  $\omega_2^{\mathcal{H}(n_\perp+n_\|)}$ & $\omega_3^{\mathcal{H}(n_\perp+n_\|)}$ \\
 $\mathcal{B}$ & $u_\perp^{\mathcal B}$ &  $u_\times^{\mathcal B}$ & $u_\|^{\mathcal B}$ &  $\omega_1^{\mathcal B}$ &
$\omega_2^{\mathcal B}$ & $\omega_3^{\mathcal B}$ \\
 ${\mathcal U}_1$ & $u_\perp^{{\mathcal U}_1}$ &  $u_\times^{{\mathcal U}_1}$ & $u_\|^{{\mathcal U}_1}$ &
$\omega_1^{{\mathcal U}_1}$ & $\omega_2^{{\mathcal U}_1}$ & $\omega_3^{{\mathcal U}_1}$ \\
 ${\mathcal U}_2$ & $u_\perp^{{\mathcal U}_2}$ &  $u_\times^{{\mathcal U}_2}$ & $u_\|^{{\mathcal U}_2}$ &
$\omega_1^{{\mathcal U}_2}$ & $\omega_2^{{\mathcal U}_2}$ & $\omega_3^{{\mathcal U}_2}$ \\
 \hline \hline
\end{tabular}
\caption{Fixed points and stability exponents of the $O(n_\|)\oplus
O(n_\perp)$ model. \label{tab1}}
\end{table}

\end{widetext}

\section{Fixed points and their stability ('phase diagram')} \label{fixp}

The FPs of the flow equations Eq.(\ref{duadl}) are given by the
solutions  of system of equations:
\begin{equation}
\label{fp} \beta_{u_a}(\{u\})=0.
\end{equation}
At the one loop order, Eq.(\ref{fp}) defines  eight fixed points.
Six of them are real and two are complex apart from the region where
$n_{\|}$ and $n_{\perp}$ are small (see table \ref{tab1}). They all
have the property to be proportional to $\varepsilon=4-d$. To
proceed with higher loop approximations one can make use of
different calculation schemes, either performing an
$\varepsilon$-expansion or solving the flow equations directly for
fixed $d=3$ (i.e. for $\varepsilon=1$) \cite{Schloms}. A particular
feature of the $\varepsilon$-expansion is that an increase of the
order of approximation does not lead to an increase of the number of
FPs.\cite{note1} Once the FPs are found in the first order of
$\varepsilon $, the next orders of the expansion only give the next
order contributions to the first order values of the FPs but do not
lead to the appearance of new FPs. On contrary, when one directly
solves non-linear flow equations for fixed $d$ in higher-loop order,
more and more fixed points may appear in addition since the order of
the polynomials to be solved for the fixed points increases.
Moreover, the physical FPs found in $\varepsilon$-expansion may
disappear when one naively solves the flow equations for fixed space
dimension $d$. It is well established by now that the expansions
involved for the field theoretic renormalization group functions are
asymptotic at best \cite{rgbooks} and one has to use appropriate
resummation procedures to get reliable results on their basis. Let
us note however that whereas in the $\varepsilon$-expansion the
resummation procedure is needed only to precise the FP values,
whereas in the fixed $d$ approach it is the resummation that allows
to judge about presence of the FP at all and comparative analysis of
the two approaches allows to judge about the FP picture on a sound
basis \cite{fixed_d}. Below we will make use of these two
approaches. We will work within the two-loop approximation
 and show that the two loop $\varepsilon$-expansion is
not sufficient even when one uses resummation in contrast to the
fixed dimension procedure.

Defining the FP picture we are looking for answers to the questions:
(i) is the critical behavior described by a certain FP with
corresponding  {\bf asymptotic} and {\bf universal} critical
exponents?  (ii) is the system very near a stability border line and
slow transients leading to an {\bf effective} and {\bf nonuniversal}
critical behavior? In such a case then the whole information
contained in the nonlinear flow equations is necessary and available
only in the second method presented here. Furthermore, in answering
these questions we demonstrate that the {\it physically} relevant
features are obtained already using the two loop approximation.

\subsection{Results in two loop $\varepsilon$-expansion}

Starting from the Eqs. (\ref{betauperp2l})--(\ref{betaupara2l}) one
gets six real FPs, as shown in the table \ref{tab1}. Four fixed
points correspond to the decoupled effective Hamiltonians of
$O(n_\perp)$ and  $O(n_\|)$ models. Besides the Gaussian FP
${\mathcal G}$($u_\perp^*=u_\times^*=u_\|^*=0$) these are the
${\mathcal H}(n_\perp)$ and ${\mathcal H}(n_\|)$ FPs with
($u_\perp^*=u^{{\mathcal H}(n_\perp)}$, $u_\times^*=u_\|^*=0$) and
($u_\|^*=u^{{\mathcal H}(n_\|)}$, $u_\times^*=u_\perp^*=0$),
correspondingly, as well as the decoupling FP ${\mathcal
D}$($u_\perp^*=u^{{\mathcal H}(n_\perp)}$, $u_\times^*=0$,
$u_\|^*=u^{{\mathcal H}(n_\|)}$). Here and below by $u^{{\mathcal
H}(n)}$ we denote the Heisenberg FP of the $O(n)$-symmetrical model.
Two more FPs correspond to the non-zero value of the coupling
$u_\times^*$. Following the nomenclature of Ref.\cite{konefi76} we
call them the isotropic Heisenberg and biconical FPs, ${\mathcal
H}(n_\perp+n_\|)$  and ${\mathcal B}$ correspondingly (see table
\ref{tab1}). In the minimal subtraction RG scheme, the value of
$u^{{\mathcal H}(n)}$ is currently known with the $\varepsilon^5$
order.\cite{Kleinert} From any of the $\beta$-functions
(\ref{betauperp2l}), (\ref{betaupara2l}) one recovers familiar
$\varepsilon^2$ result:
\begin{equation}\label{un}
u^{{\mathcal H}(n)}= \frac{6 \varepsilon}{n+8} + \frac{18
(3n+14)\varepsilon^2}{(n+8)^3}.
\end{equation}
Expressions for the coordinates of the FP ${\mathcal B}$ are too
cumbersome to be given in a compact form for general $n_{\|}$ and
$n_{\perp}$. Below we list the non zero values for the FP according
to Tab. \ref{tab1} for the physically important case $n_\|=1$,
$n_\perp=2$ we are interested in. Note, that for this case the two
FPs ${\mathcal U}_1$ and ${\mathcal U}_2$ attain complex values:
\begin{eqnarray} \begin{array}{l}
u_\|^{{\mathcal H}(1)}= 0.66667 \varepsilon + 0.41975\varepsilon^2 , \\
  u_\perp^{{\mathcal H}(2)}= 0.60000 \varepsilon + 0.36000 \varepsilon^2 , \\
  u_\|^{{\mathcal H}(3)}=0.54545 \varepsilon + 0.31104 \varepsilon^2 ,  \\
 u_\|^{{\mathcal B}}=0.40496 \varepsilon + 0.38691 \varepsilon^2 ,  \\
 u_\perp^{{\mathcal B}}=0.50569 \varepsilon + 0.31610 \varepsilon^2 ,  \\
 u_\times^{{\mathcal B}}=0.69059 \varepsilon + 0.29926 \varepsilon^2 ,  \\
  u_\|^{{\mathcal U}_1}=(0.41521 -i\ 0.32936 )\varepsilon +
 (0.21259 -i\ 0.21159 ) \varepsilon^2 ,  \\
  u_\perp^{{\mathcal U}_1}=(0.20213 + i\ 0.12400 ) \varepsilon +
 (0.15182 + i\ 0.7087 ) \varepsilon^2 ,   \\
  u_\times^{{\mathcal U}_1}=(0.98646 + i\ 0.12302 ) \varepsilon
  + (0.62025 +i\ 0.06112 ) \varepsilon^2 \end{array}  \nonumber
  \end{eqnarray}
The bad convergence of the $\varepsilon$-expansion is already shown
in the two loop order values of the biconical FP ${\mathcal B}$ in
$d=3$. This FP does not fulfill the criterion
\begin{equation}
\Delta^\star=u_\|^\star u_\perp^\star-u_\times^{\star2}>0
\end{equation}
for describing tetracritical behavior as in zero loop order (see Eq.
(5.12)  in Ref.\cite{konefi76} and the discussion in Sec. \ref{flow}
below).

From the structure of the $\beta$-functions one can derive exact
values of some of the stability exponents, as shown in the table
\ref{tab1}. There, by $\omega^{{\mathcal H}(n)}$ we denote usual
stability exponent of the $O(n)$ model. The rest of the exponents
are defined in the appropriate FPs by:
 \begin{eqnarray}\label{omegas1}
 \omega_2^{{\mathcal H}(n_\perp)}&=&\partial \beta_{u_\times} / \partial
u_\times |_{{\mathcal H}(n_\perp)}, \\ \label{omegas2}
 \omega_2^{{\mathcal H}(n_\|)} &=& \partial \beta_{u_\times} / \partial u_\times
   |_{{\mathcal H}(n_\|)}, \\ \label{omegas3}
 \omega_2^{\mathcal D}&=&\partial \beta_{u_\times} / \partial u_\times
   |_{\mathcal D}.
 \end{eqnarray}
To find the stability exponents in the FPs ${\mathcal
H}(n_\perp+n_\|)$ and ${\mathcal B}$ one should solve the
appropriate secular equation.

 \begin{table*}
\centering \tabcolsep=5mm
\begin{tabular}{lllllll}
 \hline \hline
   FP & $u_\perp^\star$ & $u_\times^\star$ & $u_\|^\star$ & $\omega_1$ & $\omega_2$ & $\omega_3$  \\
    \hline
   ${\mathcal G}$ & 0 & 0 & 0 & $-1$ & $-1$ & $-1$   \\
   $\mathcal{H}(2)$ & $1.141$ & 0 & 0 & $0.581$ & $-0.461$ & $-1$  \\
$\mathcal{H}(1)$ & 0 &  0 & $1.315$ &  $-1$ & $-0.552$ & $0.565$      \\
    $\mathcal{D}$ & $1.141$ &  0 & $1.315$ &  $0.581$ & $-0.014$ & $0.566$ \\
    $\mathcal{H}(3)$ & $1.002$ &  $1.002$ & $1.002$ &  $0.597$ &
$0.407$ & $-0.036$ \\
 $\mathcal{B}$ & $1.128$ &  $0.301$ & $1.287$ &  $0.583$ &
$0.554$ & $0.01$ \\  \hline \hline
\end{tabular}
\caption{Fixed points and stability exponents of the $O(1)\oplus
O(2)$ model obtained by the Pad\'e-Borel resummation within two
loops. Biconical FP $\mathcal{B}$ is stable. \label{tab2}}
\end{table*}

One can see from the table that three FPs ${\mathcal G}$, ${\mathcal
H}(n_\perp)$, and ${\mathcal H}(n_\|)$ are never stable for $d<4$.
However as we will see below, the stability of the other three FPs
for $d<4$ depends on the values of $n_\perp$ and $n_\|$. As far as
two stability exponents of the decoupling FP ${\mathcal D}$  are
always positive, $\omega^{{\mathcal H}(n_\perp)}, \omega^{{\mathcal
H}(n_\|)} >0$,  the stability of ${\mathcal D}$ is defined by the
sign of the exponent $\omega_2^{\mathcal D}$. Therefore, the
equation for the marginal dimensions line in the
$n_\perp-n_\|$-plane reads:
\begin{equation} \label{nD}
\partial \beta_{u_\times} / \partial u_\times
   |_{\mathcal D)} =0.
 \end{equation}
Equation (\ref{nD}) defines a curve $n_\perp^{\mathcal D}(n_\|)$
(or, equivalently $n_\|^{\mathcal D}(n_\perp)$) that borders  a
region of $n_\perp,n_\|$ values where the FP ${\mathcal D}$ is
stable. Substituting the second-order result (\ref{un}) into the
function (\ref{betautimes2l}) we get from (\ref{nD}):
\begin{equation} \label{nDeps1}
n_\perp^{\mathcal D}(n_\|) = \frac{2(16-n_\|)}{n_\|+2} -
\frac{48\varepsilon}{n_\|+2}.
 \end{equation}
Eq. (\ref{nDeps1}) can be inverted and one gets:
\begin{equation} \label{nDeps2}
n_\|^{\mathcal D}(n_\perp) = \frac{2(16-n_\perp)}{n_\perp+2} -
\frac{48\varepsilon}{n_\perp+2}.
 \end{equation}
The first term in (\ref{nDeps1}), (\ref{nDeps2}) coincides with the
first order result of Ref.\cite{konefi76} whereas the second term
gives the second-order contribution, which again demonstrates the
weakness of the second order $\varepsilon$-expansion, since the
shift to smaller values of $n_\|(n_\perp)$ in $d=3$ is drastically
overestimated leading to instability for small values of $n_\|$
and/or $n_\perp$.

Note, that the stability properties of the FP ${\mathcal D}$ can be
evaluated on the base of exact scaling arguments.\cite{scaling} At
this FP the coupling term $u_{\times}\phi_{\perp}^2 \phi_{\|}^2$
has a scaling of the product of two energy-like operators, the
latter  having scaling dimensions
$(1-\alpha_{n_\perp})/\nu_{n_\perp}$ and
$(1-\alpha_{n_\|})/\nu_{n_\|}$ correspondingly (with $\alpha_n$,
$\nu_n$ being the heat capacity and the correlation length critical
exponents in $O(n)$ universality class). In turn, this leads to the
following formula for the RG dimension $y_{u_\times}$ of the
combined operator:\cite{scaling,capevi03}
 \begin{equation}\label{scal}
y_{u_\times}= 1/\nu_{n_\perp} + 1/\nu_{n_\|}  - d.
\end{equation}
With available five-loop $\varepsilon$-expansion results for the
exponents of $O(n)$ theory\cite{Kleinert} the marginal dimensions
$n_\perp^{\mathcal D}(n_\|)$ ($n_\|^{\mathcal D}(n_\perp)$) can be
estimated.

The first order result for $n_\perp^{\mathcal D}(n_\|)$ is shown in
the figure \ref{phasediagborel} by a  dashed line. FP ${\mathcal D}$
is stable for the values of $n_\perp,n_\|$ above this line. Crossing
the line, one gets into the region where the biconical FP ${\mathcal
B}$ acquires stability. With further change of $n_\perp,n_\|$ its
coordinates do change as well and for certain value of
$n_\perp^{\mathcal H}(n_\|)$ this FP coincides with the Heisenberg
FP  ${\mathcal H}(n_\|+n_\perp)$. Then it looses its stability and
further the FP ${\mathcal H}(n_\|+n_\perp)$ is stable. Therefore,
the marginal dimension line $n_\perp^{\mathcal H}(n_\|)$ (or
$n_\|^{\mathcal H}(n_\perp)$ equivalently) can be defined from the
any of the conditions:
\begin{eqnarray} \nonumber
\beta_{u_\perp}(u_\perp,u_\perp,u_\perp)=\beta_{u_\times}(u_\times,u_\times,u_\times)=
 \\ \label{nH}
\beta_{u_\|}(u_\|,u_\|,u_\|)=0.
 \end{eqnarray}
As far as in the Heisenberg FP ${\mathcal H}$ the RG functions
depend on the sum of field dimensionalities $n_\perp+n_\|$, the
resulting marginal dimension curve will be of the form
$n_\perp^{\mathcal H}(n_\|)=const - n_\|$. Substituting  the FP
${\mathcal H}$ coordinates into expressions for the
$\beta$-functions (\ref{betauperp2l})--(\ref{betaupara2l}) we get
the second order expression for the marginal dimension. By it we
recover the two first terms of the third order result quoted in
Ref.\cite{konefi76} (and obtained from Ref.\cite{Ketley73}):
\begin{equation} \label{nHeps1}
n_\perp^{\mathcal H}(n_\|) = - n_\| + 4 - 2 \varepsilon + c^\times
\varepsilon^2,
 \end{equation}
with $c^\times=\frac{5}{12}(6\zeta(3)-1)=2.5885$. Or, equivalently:
\begin{equation} \label{nHeps2}
n_\|^{\mathcal H}(n_\perp) = - n_\perp + 4 - 2 \varepsilon +
c^\times \varepsilon^2.
 \end{equation}
In Fig. \ref{phasediagborel} the results obtained in the
$\varepsilon$-expansion in first loop order are shown. Whereas the
${\cal HB}$-stability borderlines lead to an acceptable result, but
bad convergence, the $\cal{BD}$-stability borderline show unphysical
features in the $\varepsilon$-expansion. In second order in
$\varepsilon$ the stability borderline for positive values of
$n_\perp$ lies at negative values of $n_\|$ meaning that the
decoupling fixed point is stable in the whole region shown.  The
second order $\varepsilon$ expansion results do not lead to reliable
results being estimated naively. Therefore, below we will reanalyze
the RG functions by resuming them in two loops directly for
$\varepsilon=1$.

The FPs where the parallel and perpendicular system decouples
($u_\times^\star=0$) need some comments. The renormalization group
procedure used here assumes that the multicritical system is
described by {\bf one} diverging length scale and therefore by one
correlation length $\xi$  and one corresponding critical exponent
$\nu$. This does not hold for  decoupled systems where {\bf two}
length scales and therefore two correlation lengths, $\xi_\|$ and
$\xi_\perp$ with two different asymptotic exponents $\nu_\|$ and
$\nu_\perp$ are present. Thus the usual scaling laws with one length
scale break down (see the remarks in Ref.\cite{konefi76} and Ref.
\cite{amitgoldschm78}).

\subsection{Results from resummation of the two loop field theoretic functions}

\begin{figure}[h,t,b]
      \centering{
       \epsfig{file=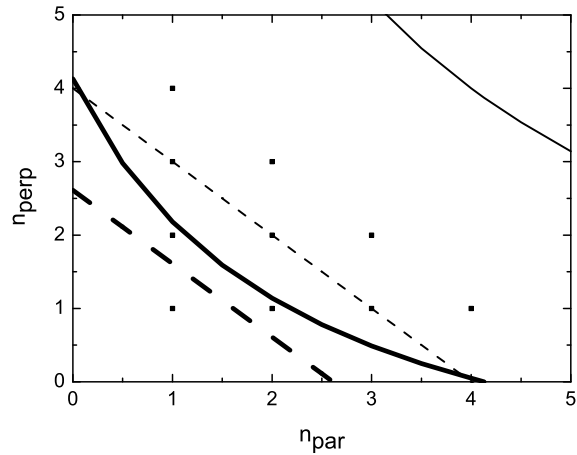,width=9cm,angle=0} }
     \caption{ \label{phasediagborel} Regions of different static bicritical behavior
in the $n_\|-n_\perp$-plane ($\varepsilon=4-d=1$), which are defined
by the stable FP (Heisenberg FP ${\mathcal H}$, biconical FP
${\mathcal B}$ and decoupling FP ${\mathcal D}$). Shown are the
${\cal HB}$-stability borderline (dashed lines) and ${\cal
BD}$-stability borderlines (solid lines), in   one loop order (thin
lines) and resummed two loop order (thick lines). The dots indicate
low integer values for OP components. One sees that the borderlines
are drastically shifted to smaller values of OP components. Thus in
the case  $n_|=1$ and $n_\perp=2$  FP ${\mathcal B}$ (connected with
tetracriticality) is stable in two loop order contrary to the one
loop order calculation where the FP ${\mathcal H}$ (connected with
bicriticality) is stable.}
\end{figure}

Let us pass now to another way of analysis of the RG functions
(\ref{betauperp2l})--(\ref{betaupara2l}). Namely, in the spirit of
the fixed dimension RG approach\cite{Schloms} let us consider the FP
equations (\ref{fp}) directly at fixed $d=3$. The RG series being
divergent, we present the $\beta$-functions
(\ref{betauperp2l})--(\ref{betaupara2l}) in a form of resolvent
series and resum them by the Pad\'e-Borel resummation technique as
explained in the Appendix \ref{resummation}. Let us note that such a
representation preserves symmetry properties of the functions. Now,
the FP coordinates as well as the marginal dimension lines are
evaluated numerically. The lines $n_\perp^{\mathcal H}(n_\|)$,
$n_\perp^{\mathcal D}(n_\|)$ are shown in Fig. \ref{phasediagborel}
by solid lines. The FP coordinates for $n_\perp=1, n_\|=2$ and the
stability exponents are given in Table \ref{tab2}.

\begin{figure}[h,t,b]
      \centering{
       \epsfig{file=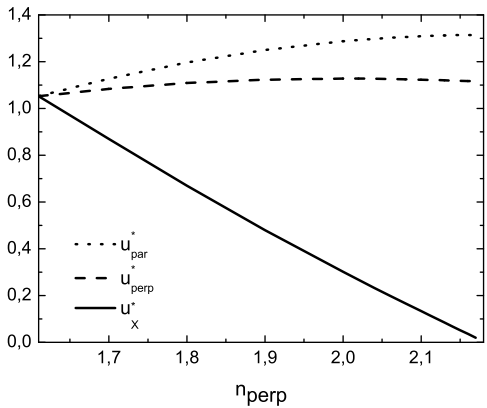,width=8cm,angle=0} }
     \caption{ \label{fixpborel} Dependence of fixed point values of the static couplings
$u_\perp$, $u_\|$, $u_\times$ (in Borel summed two loop order)  on
$n_\perp$ for $n_\|=1$ in the  region of stability of the bicritical
fixed point in $d=3$.}
\end{figure}

From the resummation at $n_\|=1$ and different $n_\perp$ we can
follow the changes in the FP values of the fourth order couplings at
the biconical FP (see Fig. \ref{fixpborel}). They start with equal
values corresponding to the isotropic Heisenberg FP at the
borderline value $n=1.6$. The cross coupling between the parallel
and perpendicular components decreases to zero at the stability
borderline to the decoupling FP at $n_{\perp}(n_\|=1)=2.17$, whereas
the other couplings slightly increase.

\section{Identification of the critical exponents} \label{expon}

The connection between critical exponents and the $\zeta$-functions can be
obtained from the solution of the renormalization group equations for the
vertex functions.

\subsection{Anomalous dimensions $\eta_\|$ and $\eta_\perp$}

Considering the vertex functions (\ref{vertexfunc3}) at the
bicritical point $\xi_\perp^{-2}=\xi_\|^{-2}=0$ the renormalization group equation
for the $k$-dependent functions read
\begin{eqnarray} \label{rngk}
\left[\kappa\frac{\partial}{\partial\kappa}+\sum_{a=\perp,\|,\times}\beta_{u_a}
\frac{\partial}{\partial u_a}+\frac{1}{2}\sum_{j=1}^{N}\zeta_{\phi_{\alpha_{j}}}
\right]  \nonumber\\
\times\quad
\Gamma^{(N,0)}_{\alpha_1\cdots\alpha_N}(k,\{u\},\kappa)=0 \, .
\end{eqnarray}
Solving the equation with the method of characteristic equations leads to
\begin{eqnarray} \label{solverngkN}
\Gamma^{(N,0)}_{\alpha_1\cdots\alpha_N}(k,\{u\},\kappa)=(\kappa l)^{\delta_N}
\exp\Big[\frac{1}{2}\sum_{j=1}^{N}\int_1^l\frac{dx}{x}\ \zeta_{\phi_{\alpha_{j}}}\Big]
\nonumber\\
\times\quad \hat{\Gamma}^{(N,0)}_{\alpha_1\cdots\alpha_N}\Big(\frac{k}{\kappa l},\{u(l)\}
\Big)
\end{eqnarray}
where $\delta_N=N+d-\frac{1}{2}Nd$ is the naive dimension of the
vertex functions. The couplings $\{u(l)\}$ are determined by the
flow equations (\ref{duadl}). For $N=2$ the vertex functions
represents the $k$-dependent inverse susceptibilities in the two
subspaces. Eq.(\ref{solverngkN}) reduces to
\begin{eqnarray} \label{solverngk2}
\Gamma^{(2,0)}_{\alpha_i\alpha_i}(k,\{u\},\kappa)=(\kappa l)^2
\exp\Big[\int_1^l\frac{dx}{x}\ \zeta_{\phi_{\alpha_{i}}}\Big]
\nonumber\\
\times\quad \hat{\Gamma}^{(2,0)}_{\alpha_i\alpha_i}\Big(\frac{k}{\kappa l},\{u(l)\}
\Big)
\end{eqnarray}
with the indices (not to confuse with the specific heat exponents below) $\alpha_i$ equal to $\perp$ or $\|$. In the
asymptotic region the couplings
$\{u(l)\}$ have nearly reached their fixed point values $\{u^\star\}$. The
$\zeta$-function $\zeta_{\phi_{\alpha_{i}}}(\{u(x)\})$ in
the exponential factor reduces to the constant $\zeta_{\phi_{\alpha_{i}}}(\{u^\star\})\equiv
\zeta_{\phi_{\alpha_{i}}}^\star$. Expression (\ref{solverngk2}) reduces to
\begin{eqnarray} \label{rngk2us}
\Gamma^{(2,0)}_{\alpha_i\alpha_i}(k,\{u\},\kappa)=\kappa^2
l^{2+\zeta_{\phi_{\alpha_{i}}}^\star}
\hat{\Gamma}^{(2,0)}_{\alpha_i\alpha_i}\Big(\frac{k}{\kappa
l},\{u^\star\}\Big) \, .
\end{eqnarray}
In order to obtain a finite amplitude function
$\hat{\Gamma}^{(2,0)}_{\alpha_i\alpha_i}$ we choose the matching condition
\begin{equation}\label{matchingk}
\frac{k}{\kappa l}=1
\end{equation}
which determines the flow parameter $l(k)$. Inserting into (\ref{rngk2us}) leads
to
\begin{eqnarray} \label{rngk2as}
\Gamma^{(2,0)}_{\alpha_i\alpha_i}(k,\{u\},\kappa)=\kappa^2
\left(\frac{k}{\kappa}\right)^{2+\zeta_{\phi_{\alpha_{i}}}^\star}
\hat{\Gamma}^{(2,0)}_{\alpha_i\alpha_i}(1,\{u^\star\}) \nonumber \\
\sim k^{2+\zeta_{\phi_{\alpha_{i}}}^\star} \, .
\end{eqnarray}
Therefore the asymptotic behavior of the inverse susceptibilities
$\chi^{-1}_{\alpha_i\alpha_i}=\Gamma^{(2,0)}_{\alpha_i\alpha_i}$ is
\begin{equation}\label{chias}
\chi^{-1}_{\alpha_i\alpha_i}\sim
k^{2+\zeta_{\phi_{\alpha_{i}}}^\star} \sim k^{2-\eta_{\alpha_{i}}}
\, .
\end{equation}
Thus we may identify the two anomalous dimensions
\begin{equation}\label{etaid}
\eta_{\alpha_{i}}=-\zeta_{\phi_{\alpha_{i}}}^\star
\end{equation}
with $\alpha_i=\perp,\|$.

\subsection{Exponents of the susceptibilities $\gamma_\|$ and $\gamma_\perp$}

In order to obtain the exponents $\gamma_\|$ and $\gamma_\perp$ and
their relation to the exponent of the correlation length $\nu$, we
have to consider the temperature dependent vertex functions and the
corresponding renormalization group equations. There are two methods
to include the temperature dependence into the renormalization group
equations. The first one is to constitute renormalization group
equations for the vertex functions (\ref{vertexfunc2}) and derive
from the renormalization of the temperature distance
$\bigtriangleup\vec{r}$ a relation between the exponents $\gamma_\|$
and $\gamma_\perp$ and the corresponding function
$\mbox{\boldmath$\zeta$}_r$. The second one is to constitute a
renormalization group equation for the vertex functions
(\ref{vertexfunc3}) including the temperature dependence by an
expansion in $\phi^2$-insertions. This would lead to a relation of
the exponents $\gamma_\|$ and $\gamma_\perp$ to the function
$\mbox{\boldmath$\zeta$}_{\phi^2}$. As a consequence of relation
(\ref{zrzphi2rel}) one obtains  the scaling laws between the
exponents $\gamma_\|$ and $\gamma_\perp$  and $\eta_\|$ and
$\eta_\perp$ with including only {\bf one} exponent $\nu$ for both
correlation length (see below). In the following we will consider
the first method. The renormalization group equations for the vertex
functions (\ref{vertexfunc2}) at $k=0$ read
\begin{eqnarray} \label{rngr}
\left[\kappa\frac{\partial}{\partial\kappa}+\sum_{a=\perp,\|,\times}\beta_{u_a}
\frac{\partial}{\partial u_a}+\bigtriangleup\vec{r}\cdot\mbox{\boldmath$\zeta$}_{\phi^2}
\cdot\frac{\partial}{\partial\bigtriangleup\vec{r}}   \right. \nonumber\\
\left.+\frac{1}{2}\sum_{j=1}^{N}\zeta_{\phi_{\alpha_{j}}}
\right]
\Gamma^{(N,0)}_{\alpha_1\cdots\alpha_N}(\{\bigtriangleup r\},\{u\},\kappa)=0
\end{eqnarray}
where relation (\ref{zetarel}) already has been used. The matrix
$\mbox{\boldmath$\zeta$}_{\phi^2}$ can be diagonalized by the transformation
\begin{equation}\label{zetatransform}
\mbox{diag}(\zeta_+,\zeta_-)=\mbox{\boldmath$P$}^{-1}\mbox{\boldmath$\zeta$}_{\phi^2}^T
\mbox{\boldmath$P$}
\end{equation}
where $\zeta_+$ and $\zeta_-$ are the eigenvalues of
$\mbox{\boldmath$\zeta$}_{\phi^2}$, while the matrix
$\mbox{\boldmath$P$}$ is determined by the corresponding
eigenvectors. The matrix $\mbox{\boldmath$\zeta$}_{\phi^2}^T$ has
the form
\begin{equation}\label{zetaphi2str}
\mbox{\boldmath$\zeta$}_{\phi^2}^T=\frac{1}{6}\left(\begin{array}{cc}
V_\perp & n_\|u_\times C \\ n_\perp u_\times C &
V_\|\end{array}\right) \, .
\end{equation}
The functions $V_\perp=V_\perp(\{u\})$, $V_\|=V_\|(\{u\})$ and
$C=C(\{u\})$ may contain all orders of loop expansion. The two loop
expressions have been given in (\ref{zetaps11})-(\ref{cu}). The
eigenvalues of this matrix read
\begin{equation}\label{eigenvalues}
\zeta_\pm=\frac{1}{12}(\Sigma\pm W)
\end{equation}
where we have introduced the square root
\begin{equation}\label{squareroot}
W=\sqrt{\Delta^2+4n_\perp n_\|u_\times^2C^2}
\end{equation}
and the parameters
\begin{equation}\label{pardef}
\Sigma=V_\perp+V_\|  \ , \qquad  \Delta=V_\perp-V_\| \, .
\end{equation}
The corresponding eigenvalues constitute the transformation matrix
\begin{equation}\label{pmatrix}
\mbox{\boldmath$P$}=\left(\begin{array}{cc}
1 & -\frac{2n_\| u_\times C}{\Delta+W} \\ \frac{2n_\perp u_\times C}{\Delta+W} &
1 \end{array}\right)
\end{equation}
where the first and the second column are the eigenvectors to
$\zeta_+$ and $\zeta_-$. The explicit appearance of the parameters
(\ref{squareroot}) and (\ref{pardef}) in the matrix (\ref{pmatrix})
may differ by using the relation
\begin{equation}\label{dwrel}
(\Delta+W)(\Delta-W)=-4n_\perp n_\|u_\times^2C^2 \, .
\end{equation}
For special cases the eigenvalues and the eigenvectors may simplify
considerably.

i) In the case of the decoupled system, where $u_\times=0$, a decay
into two isolated isotropic subsystems with order parameter
dimensions $n_\perp$ and $n_\|$ occur. The matrix (\ref{pmatrix})
reduces to the unit matrix and the eigenvalues
$\zeta_+=V_\perp/6=\zeta_{\phi^2}^{(n_\perp)}$ and $\zeta_-=V_\|/6
=\zeta_{\phi^2}^{(n_\|)}$ are the corresponding isotropic
$\zeta$-functions of the two subsystems.

ii) In the case of the isotropic Heisenberg system, i.e. $u_\perp=u_\|=u_\times=u$, the
matrix (\ref{pmatrix}) reduces to
\begin{equation}\label{pmatrixhi}
\mbox{\boldmath$P$}=\left(\begin{array}{cc}
1 & -\frac{n_\|}{n_\perp} \\ 1 & 1 \end{array}\right)
\end{equation}
which is independent on $u$ also in higher order perturbation
expansion. The first eigenvector $\vec{g}_+^T=(1,1)$, which
corresponds to $\zeta_+$, points in a $45$ degree direction  from
the bicritical point in the $\bigtriangleup r_\perp$
-$\bigtriangleup r_\|$-plane, while the second eigenvector
$\vec{g}_-^T=(-n_\|/n_\perp,1)$ lies in some sense tangentially to
it. The eigenvalues in two loop order are
\begin{eqnarray}
\label{zetapl2l}
\zeta_+&=&\frac{n_\perp+n_\|+2}{6}\ u\left(1-\frac{5}{12}u\right)
=\zeta_{\phi^2}^{(n_\perp+n_\|)}  \, , \\
\label{zetami2l}
\zeta_-&=&\frac{u}{3}\left(1-\frac{n_\perp+n_\|+10}{24}\ u\right) \,
,
\end{eqnarray}
from which one can see that $\zeta_+$ reproduces in this case the corresponding
$\zeta$-function of an isotropic $n_\perp+n_\|$-component Heisenberg system.
The Eigenvalues and the direction of the Eigenvectors imply that obviously
$\zeta_+$ defines an exponent $\nu$, while the other Eigenvalue $\zeta_-$ is
connected with a crossover exponent.

Diagonalizing $\mbox{\boldmath$\zeta$}_{\phi^2}$ and introducing transformed
temperature distances
\begin{equation}\label{rtrans}
\vec{r}_\pm\equiv\left(\begin{array}{c}
r_+ \\ r_- \end{array}\right)=\mbox{\boldmath$P$}^{-1}\bigtriangleup \vec{r}
\end{equation}
in Eq.(\ref{rngr}) we obtain the renormalization group equation
\begin{eqnarray} \label{rngrpm}
\left[\kappa\frac{\partial}{\partial\kappa}+\sum_{a=\perp,\|,\times}\beta_{u_a}
\frac{\partial}{\partial u_a}+\zeta_+r_+\frac{\partial}{\partial r_+}
+\zeta_-r_-\frac{\partial}{\partial r_-}   \right. \nonumber\\
\left.+\frac{1}{2}\sum_{j=1}^{N}\zeta_{\phi_{\alpha_{j}}} \right]
\Gamma^{(N,0)}_{\alpha_1\cdots\alpha_N}(r_+,r_-,\{u\},\kappa)=0 \, .
\end{eqnarray}
The solution of this equation is
\begin{widetext}
\begin{eqnarray} \label{solverngrN}
\Gamma^{(N,0)}_{\alpha_1\cdots\alpha_N}(r_+,r_-,\{u\},\kappa)=(\kappa l)^{\delta_N}
\exp\Big[\frac{1}{2}\sum_{j=1}^{N}\int_1^l\frac{dx}{x}\ \zeta_{\phi_{\alpha_{j}}}\Big]
\hat{\Gamma}^{(N,0)}_{\alpha_1\cdots\alpha_N}
\Big(\frac{r_+(l)}{(\kappa l)^2},\frac{r_-(l)}{(\kappa l)^2},
\{u(l)\}\Big)
\end{eqnarray}
\end{widetext}
where the transformed temperature distances fulfill the flow equations
\begin{equation}\label{drpmdl}
l\frac{d r_\pm}{dl}=r_\pm\zeta_\pm(\{u\}) \, .
\end{equation}
Considering the solution (\ref{solverngrN}) in the asymptotic region $\{u\}=\{u^\star\}$
we obtain for $N=2$
\begin{eqnarray} \label{solverngr2}
\Gamma^{(2,0)}_{\alpha_i\alpha_i}(r_+,r_-,\{u\},\kappa)=\kappa^2
l^{2+\zeta_{\phi_{\alpha_i}}^\star}  \nonumber \\
\times\ \hat{\Gamma}^{(2,0)}_{\alpha_i\alpha_i}
\Big(\frac{r_+(l)}{(\kappa l)^2},\frac{r_-(l)}{(\kappa l)^2},
\{u^\star\}\Big) .
\end{eqnarray}
As discussed above, $r_+$ is the temperature distance to the
bicritical point, thus the matching condition
\begin{equation}\label{rmatch}
\frac{r_+(l)}{(\kappa l)^2}=1
\end{equation}
defines the flow parameter $l(r_+)$ as a function of temperature. In
the asymptotic region the solution of Eq. (\ref{drpmdl}) is
\begin{equation}\label{rpml}
r_\pm(l)=r_\pm l^{\zeta_\pm^\star} \, .
\end{equation}
From (\ref{rmatch}) we obtain therefore
\begin{equation}\label{lrp}
l=\left(\frac{r_+}{\kappa^2}\right)^{\frac{1}{2-\zeta_+^\star}} \, .
\end{equation}
Inserting into (\ref{solverngr2}) leads to the asymptotic expression
\begin{eqnarray} \label{solverngr2as}
\Gamma^{(2,0)}_{\alpha_i\alpha_i}(r_+,r_-,\{u\},\kappa)=\kappa^2
\left(\frac{r_+}{\kappa^2}\right)^{\gamma_{\alpha_i}}
\nonumber \\
\times\ \hat{\Gamma}^{(2,0)}_{\alpha_i\alpha_i}
\Big(1,\frac{\frac{r_-}{\kappa^2}}{\left(\frac{r_+}{\kappa^2}
\right)^{\phi}},\{u^\star\}\Big)
\end{eqnarray}
of the inverse order parameter susceptibility and we may identify the exponent
$\gamma_{\alpha_i}$ and the crossover exponent $\phi$ as
\begin{equation}\label{gammai}
\gamma_{\alpha_i}=
\frac{2+\zeta_{\phi_{\alpha_i}}^\star}{2-\zeta_+^\star} \ , \qquad
\phi=\frac{2-\zeta_-^\star}{2-\zeta_+^\star} \, .
\end{equation}
Introducing $\eta_{\alpha_i}$ from (\ref{etaid}) into the first
relation one can write
\begin{equation}\label{gammai2}
\gamma_{\alpha_i}= \frac{2-\eta_{\alpha_i}}{2-\zeta_+^\star} =\nu(2-\eta_{\alpha_i})
\end{equation}
which obviously leads to the exponent
\begin{equation}\label{nu}
\nu^{-1}= 2-\zeta_+^\star\equiv\nu_+^{-1}
\end{equation}
of the correlation length. Although two exponents $\eta_{\alpha_i}$
($\alpha_i=\perp,\|$), and as a consequence two exponents
$\gamma_{\alpha_i}$, have been obtained, only one exponent $\nu$
describes the multicritical behavior. This reflects the fact that
the anisotropy is present only in the order parameter component
space, but not in the coordinate space. Only a {\bf single}
diverging length scale is present in the system in our case, for  a
discussion when two length scales are present see
Ref.\cite{amitgoldschm78}. For the discussion in section
\ref{exponents} it is convenient to define the exponent
\begin{equation}\label{numinus}
\nu_-^{-1}\equiv 2-\zeta_-^\star
\end{equation}
quite analogous to (\ref{nu}). The crossover exponent in (\ref{gammai}) can then
be written as
\begin{equation}\label{crossphi}
\phi=\frac{\nu_+}{\nu_-}    \ .
\end{equation}

\subsection{Exponent of the specific heat $\alpha$}

Within the Ginzburg-Landau-Wilson model the specific heat is proportional to the
$\phi^2$-$\phi^2$-correlation function, which is the negative vertex function
$\Gamma^{(0,2)}$. Therefore we have to consider the solutions of the
renormalization group equations for $\Gamma^{(0,2)}_{;\beta_1\beta_2}$ defined in
(\ref{gamma02ren}) in order to
obtain a theoretical expression for the exponent $\alpha$. Considering the functions
$\Gamma^{(0,2)}_{;\beta_1\beta_2}$ as a symmetric matrix
\begin{equation}\label{ga02}
\mbox{\boldmath$\Gamma$}^{(0,2)}=\left(\begin{array}{cc}
\Gamma^{(0,2)}_{\perp\perp} & \Gamma^{(0,2)}_{\perp\|} \\
\Gamma^{(0,2)}_{\perp\|} & \Gamma^{(0,2)}_{\|\|}\end{array}\right)
\end{equation}
the renormalization group equation reads
\begin{eqnarray} \label{rngr02}
\left[\kappa\frac{\partial}{\partial\kappa}+\!\!\!\sum_{a=\perp,\|,\times}
\!\!\!\beta_{u_a}
\frac{\partial}{\partial u_a}+\bigtriangleup\vec{r}\cdot\mbox{\boldmath$\zeta$}_{\phi^2}
\cdot\frac{\partial}{\partial\bigtriangleup\vec{r}}-\epsilon   \right]
\mbox{\boldmath$\Gamma$}^{(0,2)}  \nonumber\\
+\mbox{\boldmath$\zeta$}_{\phi^2}\cdot\mbox{\boldmath$\Gamma$}^{(0,2)}
+\mbox{\boldmath$\Gamma$}^{(0,2)}\cdot\mbox{\boldmath$\zeta$}_{\phi^2}^T
=-\mbox{\boldmath$B$}_{\phi^2}   \nonumber \\
\end{eqnarray}
where the vertex function is taken at $k=0$, i.e.
$\mbox{\boldmath$\Gamma$}^{(0,2)}=\mbox{\boldmath$\Gamma$}^{(0,2)}
(\{\bigtriangleup r\},\{u\},\kappa)$. The term $-\epsilon$ is the
naive dimension of the vertex function and it appears explicitly
because $\mbox{\boldmath$\Gamma$}^{(0,2)}$ has been introduced as
dimensionless quantity in (\ref{gamma02ren}). Introducing the
diagonalized $\zeta$-functions (\ref{zetatransform}) and the
transformed temperature distances (\ref{rtrans}), Eq.(\ref{rngr02})
can be rewritten as
\begin{eqnarray} \label{rngr02trans}
\Bigg[\kappa\frac{\partial}{\partial\kappa}+\!\!\!\sum_{a=\perp,\|,\times}
\!\!\!\beta_{u_a}
\frac{\partial}{\partial u_a}+\zeta_+ r_+\frac{\partial}{\partial r_+}
 +\zeta_- r_-\frac{\partial}{\partial r_ -}  \nonumber\\
-\epsilon   \Bigg]\mbox{\boldmath$\Gamma$}_\pm^{(0,2)}
+2\mbox{diag}(\zeta_+,\zeta_-)\cdot\mbox{\boldmath$\Gamma$}_\pm^{(0,2)}
=-\mbox{\boldmath$B$}_{\phi^2}^{(\pm)} \, .
\end{eqnarray}
The transformed vertex functions are
\begin{equation}\label{ga02trans}
\mbox{\boldmath$\Gamma$}_\pm^{(0,2)}\equiv\left(\begin{array}{cc}
\Gamma^{(0,2)}_{++} & \Gamma^{(0,2)}_{+-} \\
\Gamma^{(0,2)}_{-+} & \Gamma^{(0,2)}_{--}\end{array}\right)=
\mbox{\boldmath$P$}^T\mbox{\boldmath$\Gamma$}^{(0,2)}\mbox{\boldmath$P$}
\, .
\end{equation}
Quite analogously is $\mbox{\boldmath$B$}_{\phi^2}^{(\pm)}=\mbox{\boldmath$P$}^T
\mbox{\boldmath$B$}_{\phi^2}\mbox{\boldmath$P$}$. Applying the matrix
(\ref{pmatrixhi}) from the isotropic case one obtains $\Gamma^{(0,2)}_{++}=
\Gamma^{(0,2)}_{\perp\perp}+2\Gamma^{(0,2)}_{\perp\|}+\Gamma^{(0,2)}_{\|\|}$,
which is the specific heat in the isotropic Heisenberg model. Thus in the present
case the specific heat is obviously proportional to $\Gamma^{(0,2)}_{++}$. The
corresponding renormalization group equation is
\begin{eqnarray} \label{rngrpp}
\Bigg[\kappa\frac{\partial}{\partial\kappa}+\!\!\!\sum_{a=\perp,\|,\times}
\!\!\!\beta_{u_a}
\frac{\partial}{\partial u_a}+\zeta_+ r_+\frac{\partial}{\partial r_+}
 +\zeta_- r_-\frac{\partial}{\partial r_ -}  \nonumber\\
+2\zeta_+-\epsilon \Bigg] \Gamma_{++}^{(0,2)}=-B_{\phi^2}^{(++)}
\end{eqnarray}
with the solution
\begin{widetext}
\begin{eqnarray} \label{solverngr02}
\Gamma^{(0,2)}_{++}(r_+,r_-,\{u\},\kappa)= e^{\int_1^l\frac{dx}{x}\
(2\zeta_+-\epsilon)} \hat{\Gamma}^{(0,2)}_{++}
\Big(\frac{r_+(l)}{(\kappa l)^2},\frac{r_-(l)}{(\kappa l)^2},
\{u(l)\}\Big)+\int_1^l\frac{dx}{x}\ B_{\phi^2}^{(++)}(\{u(l)\}) \
e^{\int_1^x\frac{dx^\prime}{x^\prime}\ (2\zeta_+-\epsilon)} \, .
\end{eqnarray}
\end{widetext}
Using (\ref{rmatch}) - (\ref{lrp}) the expression reduces in the asymptotic region
to
\begin{widetext}
\begin{eqnarray} \label{solverngr02as}
\Gamma^{(0,2)}_{++}(r_+,r_-,\{u\},\kappa)=
\left(\frac{r_+}{\kappa^2}\right)^{-\alpha}\Bigg[
\hat{\Gamma}^{(0,2)}_{++}\Big(1,\frac{r_-/\kappa^2}{(r_+/\kappa^2)^{1/\phi}},
\{u^\star\}\Big)
+\frac{B_{\phi^2}^{(++)}(\{u^\star\})}
{2\zeta_+^\star-\epsilon}\Bigg]
\end{eqnarray}
\end{widetext}
where $\alpha$ is given by
\begin{equation}\label{alphaid}
\alpha= \frac{\epsilon-2\zeta_+^\star}{2-\zeta_+^\star}
=2-\frac{d}{2-\zeta_+^\star}=2-d\nu \, .
\end{equation}
In the last equality we have used Eq.(\ref{nu}). Thus the hyperscaling relation
has been derived from the renormalization group equation for the specific heat.

\subsection{Numerical estimates for asymptotic critical exponents and scaling}
\label{exponents}

The results for marginal dimensions presented above in section
\ref{fixp} give an evidence that for the physically relevant case
$n_\|=1$, $n_\perp=2$ the biconical FP $\mathcal{B}$ is stable.
Substituting the $\varepsilon$-expansion for the biconical FP
$\mathcal{B}$ we recover the results presented in
Ref.\cite{capevi03}. Although the second order of perturbation
theory is known as an optimal truncation for the
$\varepsilon$-expansion, in our case the values for the exponents
are not reliable. Resummation of the $\varepsilon$-expansion in this
order makes not much sense.

\begin{table*}
\centering \tabcolsep=2mm
\begin{tabular}{llllllllll}
 \hline \hline
 Reference &
   FP & $\eta_\perp$ & $\eta_\|$ & $\gamma_\perp$ & $\gamma_\|$ & $\nu_+$ & $\nu_-$ & $\phi$
   & $\alpha$ \\ \hline
 this paper &   ${\mathcal B}$ & 0.037 & 0.037 & 1.366 & 1.366 &
0.696 & 0.692 & 1.144\footnotemark[7]
 \footnotetext[7]{Pole in the Pad\'e approximant is present (see Appendix \ref{resummation}).} & -0.088       \\
 &   ${\mathcal H}(3)$ & 0.040 & 0.040 & 1.411 & 1.411 & 0.720 & 0.564 & 1.275\footnotemark[7] & -0.160     \\
 Ref.\cite{nekofi74} &  ${\mathcal B}$ & 0 & 0 & 1.222 & 1.222  & 0.611 &  0.503  & 1.176 &  {\em 0.167}     \\
Ref.\cite{nekofi74} &   ${\mathcal H}(3)$ & 0 & 0 & 1.227 & 1.227 & 0.611 & 0.505 &  1.136  & {\em 0.167}       \\
Ref.\cite{capevi03} &   ${\mathcal B}$ & 0.037(5) & 0.037(5) & {\em
1.37(7)} & {\em 1.37(7)} & 0.70(3) & {\em 0.56(3)}  & 1.25(1) &
{\em -0.10(9)}     \\
Ref.\cite{Guida98} &   ${\mathcal H}(3)$ & 0.0375(45) & 0.0375(45) &
1.382(9) & 1.382(9)  & 0.7045(55) & 0.559(17) & {\em 1.259(23)}   &
{\em -0.114(17)}      \\
     \hline \hline
\end{tabular}
\caption{Critical exponents of the $O(1)\oplus O(2)$ model obtained
by resummation of the two-loop RG series at fixed $d=3$ in different
FPs (first two rows of the table). Our data is compared with the
results of Ref.\cite{nekofi74} (first order
$\varepsilon$-expansion), and Refs.\cite{capevi03,Guida98} (resummed
fifth order $\varepsilon$-expansion). Numbers, shown in italic were
obtained via familiar scaling relations.
 \label{tab3}}
\end{table*}

Therefore we choose another way to estimate the numerical values of
the exponents making use of the fixed $d$ expansions. There, we
proceed as follows. First, based on the two-loop expressions
(\ref{zetaps11})-(\ref{zetaps22}) for the matrix
$[\mbox{\boldmath$\zeta$}_{\phi^2}]_{ij}(\{u\})$ (\ref{zetaphiq}) we
find its eigenvalues as expansions in renormalized couplings. In
turn, the exponents $\nu_+$, $\nu_-$ are expressed in terms of the
corresponding eigenvalues as series in renormalized couplings as
well. Corresponding series are found also for the magnetic
susceptibility exponents. The series are then resummed (as described
in the Appendix \ref{resummation}) and evaluated at the FPs.
Numerical results of the exponents $\nu_+$, $\nu_-$, $\gamma_\|$,
and $\gamma_\perp$ obtained by such evaluation are given in table
\ref{tab3} in the (stable) biconical FP ${\mathcal B}$ and, for
comparison, in the (unstable) Heisenberg FP ${\mathcal H}$. The
expressions for the exponents $\eta_\perp$ and $\eta_\|$ are too
short to be resummed. Therefore their numerical values are found
from the familiar scaling relations $\eta_\|= 2-\gamma_\|/\nu_+$,
$\eta_\perp=2- \gamma_\perp/\nu_+$ using resummed values of
$\gamma$s and $\nu$s. For the decoupling FP one has to modify the
scaling relations according to the statements above and has to use
$\eta_\|= 2-\gamma_\|/\nu_-$, $\eta_\perp=2- \gamma_\perp/\nu_+$.
Note that within the accuracy of calculations results for the
exponents that correspond to parallel and perpendicular fields do
not differ (the difference shows up within the fourth digit). The
overall agrement with the 5 loop $\varepsilon$-expansion is very
good especially for the stable biconical FP.

\section{Flow equations and effective exponents} \label{flow}

The resummation of the $\beta$-function has the big advantage to
find the fixed point values and asymptotic exponents but in addition
the flow of the couplings from their background values to their FP
values. The flow is important for the crossover behavior but also
for the discussion of the asymptotic multicritical behavior of
physical representatives of such systems with different the
background values of the couplings. Their location in the attraction
region of a FP defines the critical behavior..

\subsection{Flow of fourth order couplings}

\begin{figure}[h,t,b]
      \centering{
       \epsfig{file=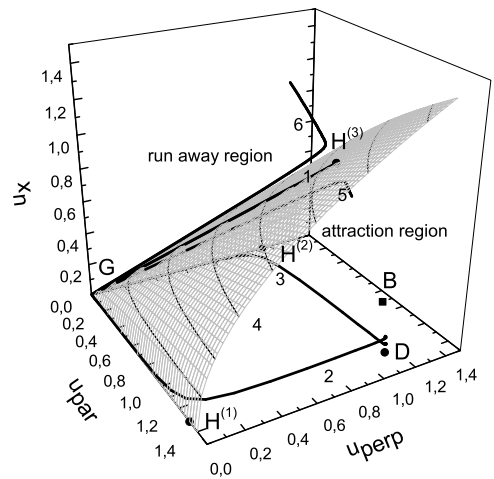,width=9cm,angle=0} }
     \caption{ \label{flowres} Resummed flow for different initial conditions. The unstable FPs are shown as filled spheres, the
stable biconical FP as filled cube. In order to show the crossover
the initial values of couplings are chosen accordingly. Due to the
small transient exponent present the stable fixed point is not
reached for the small value of the flow parameter chosen. The fixed
points are connected by separatrices defining the surface which
encloses the attraction region it is slightly different from the
MFS-surface shown (for further details see text). The initial values
of flow 6 are outside the attraction regions. The flow parameter is
changed in the interval $-40 \le \ln l \le 0$. }
\end{figure}

Fig. \ref{flowres} shows the flow lines in the space of the coupling parameters for different initial conditions
calculated from the resummed $\beta$-functions.
Mean field theory states a criterion\cite{liufi72} which has to be fulfilled by the fourth order couplings for  the existence of a tetracritical point; it reads
\begin{equation} \label{MFS}
\Delta=u_\|u_\perp-u_\times^2>0 \, .
\end{equation}
The flow equations show that $\Delta$ is not an invariant surface of the flow. However it contains several separatrices with
the corresponding fixed points: the decoupling FPs  with $u_\times^\star=0$ ( ${\mathcal G}$, ${\mathcal H(1)}$, ${\mathcal H(2)}$, ${\mathcal D}$)
and the Heisenberg FP ${\mathcal H}(3)$.  Therefore in the region shown in Fig. 1 the mean field condition $\Delta=0$
represents quite well the
surface  (called MFS) separating the attraction region of the biconical FP  ${\mathcal B}$ and the region of run away flows.

Thus one can draw the following conclusions. A system with initial
conditions $\Delta<0$ lies outside the attraction region of the
stable FB and its runaway flow indicates that a first order
transition is expected. One would conclude that the multicritical
point is a triple point. Systems with initial conditions where
roughly $\Delta=0$ would flow to the Heisenberg FP ${\mathcal H}(3)$
indicating that the multicritical point is a bicritical point. Then
finally if the initial conditions are such that $\Delta>0$ the
biconical FP ${\mathcal B}$ is reached and the multicritical point
is a tetracritical point. In this way the three scenarios sketched
in Ref. \cite{capevi03} are realized. The important and open point
is to connect the initial conditions of the field theoretic flow
equations  to the interaction parameters in the appropriate spin
Hamiltonian containing the anisotropic interactions defining the
antiferromagnetic system.

\subsection{Effective exponents}

Having available the solutions of the flow equations one can define
the effective exponents evaluating the field theoretic
$\zeta$-functions at the values of the couplings given by the flow
according to the definitions of the exponents in Eqs. (\ref{etaid})
and (\ref{nu}). We substitute the couplings obtained from the
resummed flow equations into the resummed $\zeta$-functions
appearing in the expressions for exponents. In this way the
effective exponents become functions of the flow parameter $l$, e.g.
\begin{eqnarray}
\nu_{+eff}^{-1}(l)&=& 2-\zeta_+\left(u_\|(l), u_\perp(l),u_\times(l)\right) =\nu_{eff}^{-1}(l) \nonumber \\
\nu_{-eff}^{-1}(l)&=& 2-\zeta_-\left(u_\|(l), u_\perp(l),u_\times(l)\right)
\end{eqnarray}
The effective crossover exponent $\phi_{eff}$ follows from Eq.
(\ref{crossphi}) as
\begin{equation}
\phi_{eff}(l)=\frac{\nu_{+eff}(l)}{\nu_{-eff}(l)}
\end{equation}

\begin{figure}[h,t,b]
      \centering{
              \epsfig{file=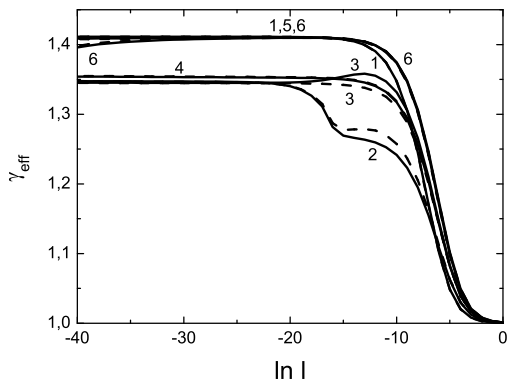,width=8cm,angle=0}}
     \caption{ \label{gammaeff} Effective exponent $\gamma_\|$ (solid lines) and $\gamma_\perp$ (dashed lines) for the
initial conditions of flows shown in Fig. \ref{flowres}}
\end{figure}

\begin{figure}[h,t,b]
      \centering{
       \epsfig{file=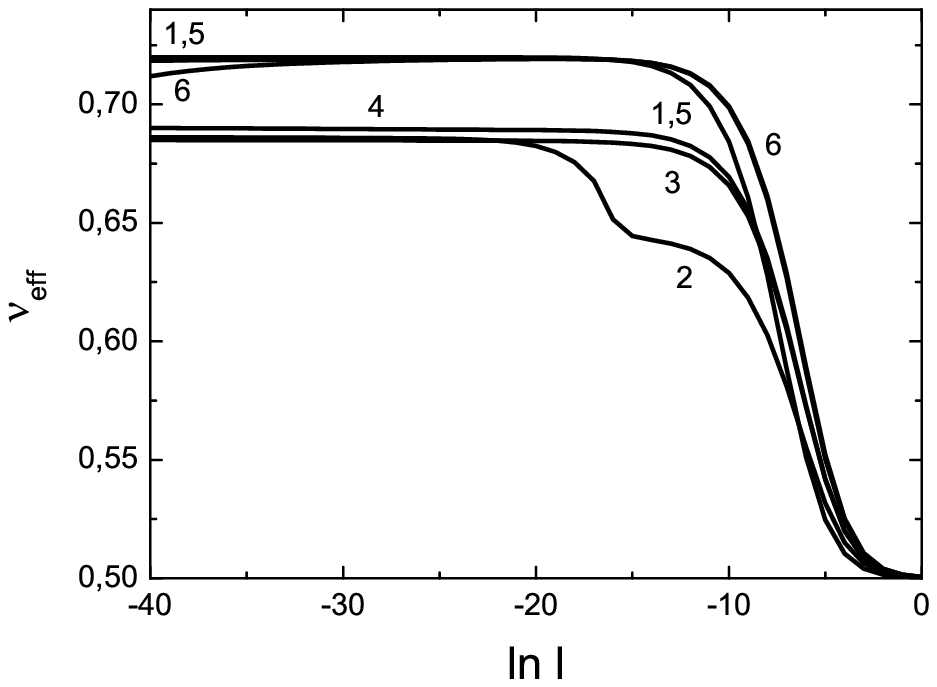,width=8cm,angle=0}}
\caption{ \label{nueff} Effective exponents $\nu$ for the
initial conditions of flows shown in Fig. \ref{flowres}}
\end{figure}

\begin{figure}[h,t,b]
      \centering{
       \epsfig{file=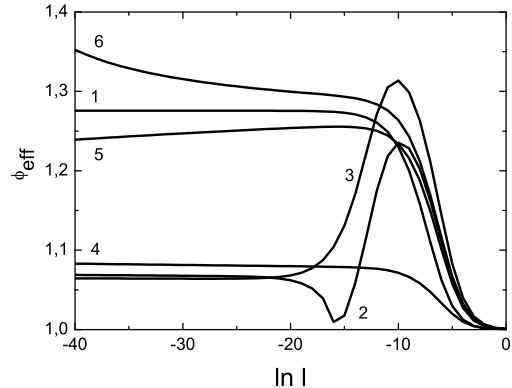,width=8cm,angle=0} }
            \caption{ \label{phieff} Effective exponents $\phi$ for the
initial conditions of flows shown in Fig. \ref{flowres}}
\end{figure}

Results for the exponents are shown in the Figs. \ref{gammaeff}, \ref{nueff} and \ref{phieff}.

The two  exponents  of the parallel and perpendicular OP
susceptibilities are almost equal even in the nonasymptotic region
but might be quite different from the asymptotic value. Especially
due to the slow transient present the values for the flows 1-3 are
smaller than the expected asymptotic values. The same holds for the
effective exponent of the correlation length $\nu_{eff}$. The flow 6
which does not reach a finite fixed point lies within the expected
values. The formal decrease  of the effective exponents to
unphysical negative values are due to the  flow to infinite values
of the couplings.

\section{Conclusion and outlook} \label{out}

We have shown that the two loop order perturbation theory together
with appropriate resummation techniques is sufficient to calculate
the multicritical behavior appearing  in systems   with
$O(n_\|)\oplus O(n_\perp)$ symmetry at fixed dimension. The
advantage of such a procedure lies in the accessibility of the
corresponding flow equations, which allow a discussion of attraction
regions and effective (crossover) critical behavior. We confirmed
the  shift of the one loop stability borderlines, with the
consequence that the multicritical behavior for the case $n_\|=1$
and $n_\perp=2$ is characterized by the stable biconical fixed point
and not by to Heisenberg fixed point. The discussion of the
attraction region of this fixed point leads to the possibility of
different phase diagrams depending on the nonuniversal initial
parameters entering the flow equations.

In a next step the results will be used to reconsider dynamical
critical behavior of the pure relaxational
dynamics\cite{dohmjanssen77} of these  systems. For the bicritical
dynamics of an antiferromagnet in an external magnetic field an
extension of the above mentioned model is necessary since besides
the OP conserved densities (CD) have to be taken into account. In
statics these couplings only appear up to second order terms and can
be integrated out. In dynamics they lead to a coupling of the two
dynamic equations. A complete description of the dynamical
multicritical behavior for has not been given in two loop order so
far. Finally mode coupling terms have to be taken into account up to
two loop order.

Acknowledgement:   We thank W. Selke for useful discussions. This
work was supported by the Fonds zur F\"orderung der
wissenschaftlichen Forschung under Project No. P19583-N20.

\appendix

\section{Vertex functions and perturbation expansion} \label{vertex}

With the static functional (\ref{hbicrit}) the vertex functions
\begin{equation}\label{vertexfunc}
\mathring{\Gamma}^{(N,L)}_{\alpha_1\cdots\alpha_N;\beta_1\cdots\beta_L}
\Big(\{\mathring{r}\},\{\mathring{u}\},k\Big)
\end{equation}
can be
calculated in a definite loop order by collecting all one-particle-irreducible
contributions. $N$ and $L$ are the total numbers of $\phi$- and $\phi^2$- insertions
and the indices $\alpha_i$ and $\beta_j$ indicate if the corresponding insertion
is of type $\perp$ or $\|$. $\{\mathring{r}\}\equiv\mathring{r}_\perp,
\mathring{r}_\|$ and $\{\mathring{u}\}\equiv\mathring{u}_\perp,
\mathring{u}_\times,\mathring{u}_\|$ act as placeholder for the two temperature distances
and three fourth order
couplings. Between the lower critical dimension $d_u=2$ and the upper critical
dimension $d_c=4$ the vertex functions (\ref{vertexfunc}) also contain
singularities at $d=3$. They have their origin in a non analytical shift of the
critical temperature as function of the four point couplings (for more details see the third refeerence in
Ref.\cite{Schloms}). n order to remove them, two parameters
$\mathring{r}_{\perp c}$ and $\mathring{r}_{\|c}$ are introduced. They are
determined by
\begin{eqnarray}
\label{r0cperp}
\mathring{\Gamma}^{(2,0)}_{\perp\perp}(\{0\},\{\mathring{u}\},0)=
\mathring{r}_{\perp c} \, , \\
\label{r0cpara}
\mathring{\Gamma}^{(2,0)}_{\|\|}(\{0\},\{\mathring{u}\},0)=\mathring{r}_{\|
c} \, .
\end{eqnarray}
This defines to functions $\mathring{r}_{\perp c}(\{\mathring{u}\})$ and
$\mathring{r}_{\| c}(\{\mathring{u}\})$. Introducing
\begin{eqnarray}
\label{deltar0perp}
\bigtriangleup\mathring{r}_\perp&=&\mathring{r}_\perp-\mathring{r}_{\perp
c}(\{\mathring{u}\}) \, ,
 \\
\label{deltar0cpara}
\bigtriangleup\mathring{r}_\|&=&\mathring{r}_\|-\mathring{r}_{\| c}(\{\mathring{u}\})
\end{eqnarray}
and rewriting the expressions for (\ref{vertexfunc}) leads to vertex functions
\begin{equation}\label{vertexfunc2}
\mathring{\Gamma}^{(N,L)}_{\alpha_1\cdots\alpha_N;\beta_1\cdots\beta_L}
\Big(\{\bigtriangleup\mathring{r}\},\{\mathring{u}\},k\Big) \, .
\end{equation}
For dimensions larger than $2$ all perturbation contributions now have only
singularities at least at $d=4$. Further
it may be convenient to introduce the correlation length instead of the temperature
distance. In the present case two correlation lengths
\begin{eqnarray}
\label{correlperp}
\xi^2_\perp(\{\bigtriangleup\mathring{r}\},\{\mathring{u}\})=
\frac{\partial}{\partial k^2}\ln\mathring{\Gamma}^{(2,0)}_{\perp\perp}
\Big(\{\bigtriangleup\mathring{r}\},\{\mathring{u}\},k\Big)\Bigg\vert_{k=0} \, , \\
\label{correlpara}
\xi^2_\|(\{\bigtriangleup\mathring{r}\},\{\mathring{u}\})=
\frac{\partial}{\partial k^2}\ln\mathring{\Gamma}^{(2,0)}_{\|\|}
\Big(\{\bigtriangleup\mathring{r}\},\{\mathring{u}\},k\Big)\Bigg\vert_{k=0}
\end{eqnarray}
have to be introduced. Inserting the reversed equations
$\bigtriangleup\mathring{r}_\perp=\bigtriangleup\mathring{r}_\perp
(\{\xi\},\{\mathring{u}\})$ and $\bigtriangleup\mathring{r}_\|
=\bigtriangleup\mathring{r}_\|(\{\xi\},\{\mathring{u}\})$ into
(\ref{vertexfunc2}) leads to functions
\begin{equation}\label{vertexfunc3}
\mathring{\Gamma}^{(N,L)}_{\alpha_1\cdots\alpha_N;\beta_1\cdots\beta_L}
\Big(\{\xi\},\{\mathring{u}\},k\Big) \, .
\end{equation}
Introducing the correlation lengths in
the vertex functions is a resummation procedure which removes the expanded
contributions of the correlation length from the vertex functions. This leads
to expressions which are simplified considerably. Moreover this is true
for the calculations in dynamic models (see for instance part II).
Within statics the two point functions reveal then the general structure
\begin{eqnarray}
\label{twopintvert}
\mathring{\Gamma}^{(2,0)}_{\perp\perp}\Big(\{\xi\},\{\mathring{u}\},k\Big)
= \frac{\mathring{f}_\perp(\{\xi\},\{\mathring{u}\})}{\xi_\perp^{2}}
+k^2\mathring{g}_\perp(\{\xi\},\{\mathring{u}\},k)\, ,
\nonumber \\
\\
\label{twopointpara}
\mathring{\Gamma}^{(2,0)}_{\|\|}\Big(\{\xi\},\{\mathring{u}\},k\Big)
= \frac{\mathring{f}_\|(\{\xi\},\{\mathring{u}\})}{\xi_\|^{2}}
+k^2\mathring{g}_\|(\{\xi\},\{\mathring{u}\},k) \, .
\nonumber \\
\end{eqnarray}
In two loop order the $k$-independent function $\mathring{f}_\perp$ is
\begin{eqnarray}
\label{falphai}
\mathring{f}_\perp(\{\xi\},\{\mathring{u}\})
=1-\frac{n_\perp+2}{18}u_\perp^2\nabla_{k^2}
D_3^{\perp\perp\perp}(\xi_\perp,0)  \nonumber \\
-\frac{n_\|}{18}u_\times^2\nabla_{k^2}
D_3^{\perp\perp\|}(\{\xi\},0)
\end{eqnarray}
where we have introduced the short notation
\begin{equation}
\label{nablaksq}
\left.\nabla_{k^2}A(\xi,0)\equiv\frac{\partial A(\xi,k)}{\partial k^2}
\right\vert_{k=0}
\end{equation}
The $k$-dependent function $\mathring{g}_\perp$ reads in two loop order
\begin{eqnarray}
\label{galphai}
\mathring{g}_\perp(\{\xi\},\{\mathring{u}\},k)=1  \nonumber \\
-\frac{n_\perp+2}{18}u_\perp^2\frac{1}{k^2}\Big(D_3^{\perp\perp\perp}(\xi_\perp,k)
-D_3^{\perp\perp\perp}(\xi_\perp,0)\Big) \nonumber \\
-\frac{n_\|}{18}u_\times^2\frac{1}{k^2}\Big(D_3^{\perp\perp\|}(\{\xi\},k)
-D_3^{\perp\perp\|}(\{\xi\},0)\Big) \, .
\end{eqnarray}
In order to obtain $\mathring{f}_\|$ and $\mathring{g}_\|$ one only has to interchange
$\perp$ and $\|$.
The two loop integral $D_3$ is defined as
\begin{widetext}
\begin{eqnarray}
\label{d3} D_3^{\alpha_i\alpha_j\alpha_k}(\{\xi\},k)=
\int_{k^\prime}\int_{k^{\prime\prime}}\frac{1}{(\xi_{\alpha_i}^{-2}+k^{\prime
2}) (\xi_{\alpha_j}^{-2}+k^{\prime\prime 2})\big(\xi_{\alpha_k}^{-2}
+(k+k^{\prime}+k^{\prime\prime})^2\big)} \, .
\end{eqnarray}
\end{widetext}
In the limit $k\to 0$ we have
\begin{eqnarray}
\label{diffquo}
\left.\lim_{k\to 0}\frac{A(\xi,k)-A(\xi,0)}{k^2}=\frac{\partial A(\xi,k)}{\partial k^2}
\right\vert_{k=0} \nonumber \\
=\nabla_{k^2}A(\xi,0)
\end{eqnarray}
Applying the limit to $\mathring{g}_{\alpha_i}$ (see (\ref{falphai}) and (\ref{galphai}))
one obtains
\begin{eqnarray}
\label{geq0}
\lim_{k\to 0}\mathring{g}_{\alpha_i}(\{\xi\},\{\mathring{u}\},k)=
\mathring{f}_{\alpha_i}(\{\xi\},\{\mathring{u}\})
\end{eqnarray}
and the two point vertex functions reduce in the asymptotic region to
\begin{eqnarray}
\label{twopintas}
\mathring{\Gamma}^{(2,0)}_{\alpha_i\alpha_i}\Big(\{\xi\},\{\mathring{u}\},k\Big)
&{\sim \atop {k\to 0}}&(\xi_{\alpha_i}^{-2}+k^2)
\mathring{f}_{\alpha_i}(\{\xi\},\{\mathring{u}\})  \nonumber \\
\end{eqnarray}
Because the poles do not depend on $k$, the two functions $\mathring{f}_{\alpha_i}$
and $\mathring{g}_{\alpha_i}$ contain the same pole terms.
Thus the two point functions each with may be renormalized by scalar renormalization factors
which remove the poles from the functions $\mathring{f}_{\alpha_i}$.

\section{Resummation}  \label{resummation}

In this appendix we describe a procedure we use to resum divergent
expansions for the two-loop RG functions. Starting from the RG
function that has a form of truncated polynomial in renormalized
couplings:
\begin{equation}\label{ap1}
f(\{u\})=\sum_{i,j,k=0}^Lc_{ijk}u_\perp^i u_\|^j u_\times^k
\end{equation}
one first represents it in a form of a resolvent
series\cite{Watson74} in variable $t$:
\begin{equation}\label{ap2}
F(t)=\sum_{i,j,k=0}^Lc_{ijk}u_\perp^i u_\|^j u_\times^k t^{i+j+k} =
\sum_{i=0}^L a_i(\{u\},\{c\})t^i.
\end{equation}
The expansion coefficients $a_i$ in (\ref{ap2}) explicitly depend on
couplings and coefficient $c_{ijk}$ (\ref{ap1}). Obviously, for
$t=1$ the function (\ref{ap1}) reproduces the initial RG function
(\ref{ap2}): $F(1)=f(\{u\})$. Then, the function (\ref{ap2}) is
resummed as a single variable function and further evaluated at
$t=1$ to recover (\ref{ap1}). To perform the resummation we use the
Pad\'e-Borel technique.\cite{Baker} Namely, assuming factorial
growth of the expansion coefficients $a_i$ we define the Borel
transform\cite{Hardy48} of (\ref{ap2}) by:
\begin{equation}\label{ap3}
F^B(t)= \sum_{i=0}^L a_i/\Gamma(i+1) t^i,
\end{equation}
where $\Gamma(x)$ is Euler gamma-function. Analytical continuation
of function (\ref{ap3}) is achieved by representing it in a form of
a Pad\'e approximant.\cite{Baker81} In our case, working within a
two-loop approximation we use the diagonal [1/1] Pad\'e approximant:
\begin{equation}\label{ap4}
F^B(t) \simeq [1/1](t).
\end{equation}
Finally, the resummed function is obtained via an inverse Borel
transform:
\begin{equation}\label{ap5}
F^{\rm res}= \int_0^{\infty} [1/1](t) e^{-t}.
\end{equation}

The procedure described above was used to analyze the
RG flows and exponents. Note however, that the inverse Borel
transform (\ref{ap5}) is well defined, when no poles in the
denominator of Pad\'e approximant (\ref{ap4}) appear. Otherwise one
may estimate its principal value. The poles do not appear for a
sign-alternating series (as the series for the $\beta$-functions
are). To deal with sign alternating series during an evaluation of
critical exponents, we have resummed the functions $2-\zeta_{\pm}$.

\end{document}